\documentclass[manuscript]{aastex}

\newcommand{\feii}{\ion{Fe}{2}}
\newcommand{\feiii}{\ion{Fe}{3}}
\newcommand{\hi}{\ion{H}{1}}
\newcommand{\hii}{\ion{H}{2}}
\newcommand{\sii}{\ion{S}{2}}
\newcommand{\siii}{\ion{S}{3}}
\newcommand{\oi}{\ion{O}{1}}
\newcommand{\oii}{\ion{O}{2}}
\newcommand{\oiii}{\ion{O}{3}}
\newcommand{\n}{\ion{N}{1}}
\newcommand{\nai}{\ion{Na}{1}}
\newcommand{\nii}{\ion{N}{2}}
\newcommand{\mgi}{\ion{Mg}{1}}

\newcommand{\caii}{\ion{Ca}{2}}
\newcommand{\crii}{\ion{Cr}{2}}

\newcommand{\coii}{\ion{Co}{2}}

\newcommand{\clii}{\ion{Cl}{2}}
\newcommand{\ariii}{\ion{Ar}{3}}
\newcommand{\ariv}{\ion{Ar}{4}}
\newcommand{\ci}{\ion{C}{1}}
\newcommand{\pii}{\ion{P}{2}}
\newcommand{\neiii}{\ion{Ne}{3}}

\newcommand{\tii}{\ion{Ti}{2}}

\newcommand{\NIii}{\ion{Ni}{2}}

\newcommand{\he}{\ion{He}{1}}
\newcommand{\heii}{\ion{He}{2}}
\newcommand{\htwo}{H$_2$}

\shorttitle{Solving the excitation and chemical abundances in shocks: the case of HH\,1}
\shortauthors{Giannini et al.}

\begin{document}


\title{Solving the excitation and chemical abundances in shocks: the case of HH\,1\thanks{Based on observations collected at the European Southern Observatory, (92.C-0058)}}

\author{T. Giannini\altaffilmark{1}, S. Antoniucci\altaffilmark{1} and B. Nisini\altaffilmark{1}}
\affil{INAF-Osservatorio Astronomico di Roma, via Frascati 33, I-00040 Monte Porzio Catone, Italy}



\and

\author{F. Bacciotti\altaffilmark{3} and L. Podio\altaffilmark{3}}
\affil{INAF-Osservatorio Astrofisico di Arcetri, Largo E. Fermi 5, I-50125 Firenze, Italy}




\begin{abstract}
We present deep spectroscopic (3\,600 - 24\,700 \AA\,) X-shooter observations of the bright Herbig-Haro object HH\,1, one of the best laboratories to study the chemical and physical modifications caused by protostellar shocks on the natal cloud.
We observe atomic fine structure lines, \hi\, and \he\, recombination lines  and \htwo\, ro-vibrational lines (more than 500 detections in total). Line emission was analyzed by means of Non Local Thermal Equilibiurm codes to derive the electron temperature and density, and, for the first time, we are able to accurately probe different physical regimes behind a dissociative shock. We find a temperature stratification in the range 4\,000 K $\div$ 80\,000 K, and a significant correlation between temperature and ionization energy. Two density regimes are identified for the ionized gas, a more tenuous, spatially broad component (density $\sim$ 10$^3$ cm$^{-3}$), and a more compact component (density $\ge$ 10$^5$ cm$^{-3}$) likely associated with the hottest gas.  A further neutral component is also evidenced, having temperature $\la$ 10\,000 K and  density $>$ 10$^4$ cm$^{-3}$. The gas fractional ionization was estimated solving the ionization equilibrium equations of atoms detected in different ionization stages. We find that neutral and fully ionized regions co-exist inside the shock. Also, indications in favor of at least partially dissociative shock as the main mechanism for molecular excitation are derived.
Chemical abundances are estimated for the majority of the detected species. On average, abundances of non-refractory/refractory elements are lower than solar of about 0.15/0.5 dex. This testifies the presence of dust inside the medium, with a depletion factor of Iron of $\sim$40\%. 

\end{abstract}






\section{Introduction}{\label{sec:sec1}

A fundamental aspect of the star formation is the evolution of the physical and chemical
properties induced by the new born star(s) in the circumstellar matter. Such interaction quite often occurs through a variety of mass ejection phenomena in the form of supersonic stellar winds and jets, which generate shock waves in the medium. These phenomena, together with the acceleration, compression, and heating of the surrounding environment, are often accompanied by the injection of ionizing UV photons and by the total or partial disruption of the dust. As a consequence, substantial modifications of the chemical structure (abundances and ionization degree) and physical conditions (temperature, density) of the natal cloud can be induced. These processes, in turn, may affect the formation of the subsequent generations of stars and can possibly lead to the dispersal of the cloud itself. 

From an observational point of view, shock waves in protostellar flows are recognizable as bright collimated knots and giant bow-shocks, where part of the kinetic energy is turned into thermal motion, and the gas is partially ionized and heated up to 10$^{5}$ K (e.g. Hollenbach \& McKee 1989). A number of forbidden lines are thus
produced, whose intensities and intensity ratios are commonly used to infer on the physical conditions in the shocked gas (e.g. Giannini et al. 2008, 2013, Maurri et al. 2014, Mesa-Delgado et al. 2009, Nisini et al. 2005, Pesenti et al. 2003, Podio et al. 2006, 2009, Hartigan et al. 1994, Hartigan \& Morse 2007). Typical tracers observable from the ground are bright lines of abundant ionic species (such S, O, N) in the UV/optical range, and H$_2$ lines in the near-infrared. Such lines are excited under different conditions of temperature and density, which span from a few thousands to more than 10$^5$ Kelvin and from 10$^2$ cm$^{-3}$ to 10$^7$ cm$^{-3}$, respectively. Observationally, a detailed mapping of the excitation conditions requires the detection of a quite large number of lines and therefore to obtain deep spectra in an extended wavelength range. Unfortunately, these requirements are rarely achieved simultaneously, and therefore the shocked regions are often described assuming oversimplified temperature and density conditions. 
Moreover, diagnostic tools based on ratios between lines of different species rely on assumptions about the relative abundances, which can be significantly different from the solar values (e.g. Podio et al. 2006 and references therein). Indeed, there are many observational evidences that solar abundances may not be representative of the local 
ISM 4.6 Gyr ago, when the Sun was formed (Wilson \& Rood 1994). On the other hand, although several observational studies have been dedicated to estimate the chemical composition in environments
different from the solar neighborhood (Grevesse \& Sauval 1998, Esteban et al. 2004, Asplund et al. 2005), we are still far from a complete view of the chemistry of shock environments, especially as far as the less abundant elements are concerned.

Another fundamental issue to investigate is the chemical modifications induced in the local dust due to the passage of shock waves. Both observations and theoretical models show that in the unperturbed medium atoms like Iron (Fe), Titanium (Ti), Magnesium (Mg), Silicon (Si), and Calcium (Ca) are considerably depleted on the dust grains, and their gas-phase abundances are 10$^2$-10$^4$ times lower than the solar abundances  (e.g. Savage \& Sembach 1996, May et al. 2000, Mouri \& Taniguchi 2000). 
However, in the high energy gas-grain and grain-grain collisions occurring behind the shock front, grain surfaces can be eroded (sputtering) and/or vaporised  (Draine 2003, Jones 2000, Jones et al. 1994, Guillet et al. 2009 and references therein), and the gas-phase abundances of the refractory elements are considerably enhanced. While some studies have been dedicated to the estimate of the depletion degree of Iron in protostellar jets (Beck-Winchatz et al. 1994, Nisini et al. 2002, Agra-Amboage et al. 2011), very little is known about the reprocessing of other species  (Nisini et al. 2005, Garcia Lopez et al. 2008,  Giannini et al. 2008 Podio et al. 2009, 2011).

In this paper we exploit the high sensitivity and the large spectral range covered by the X-shooter spectrograph at ESO-VLT to obtain an atlas as complete as possible of the ionic species involved in the chemistry of a remarkably bright protostellar bow-shock,  namely the Herbig-Haro object HH\,1. 
Being one of the brightest Herbig-Haro objects in the sky, it represents a reference frame for all the studies of HH objects (for a review of the observational and theoretical studies see Raga et al. 2011). Together with HH\,2, it was the first discovered Herbig-Haro object (Herbig 1951, Haro 1952). It is a bow-shock at the NW edge of a giant bipolar outflow almost 3 arcminutes long, corresponding to a projected length of 0.35 pc at the distance of $\sim$ 414 pc of the Orion Nebula Cluster where the object is located.
It is powered by the VLA-1 source (discovered by Pravdo 1985), a protostar roughly positioned at the center of the bipolar outflow. Spectroscopy of the HH\,1/HH\,2 system was performed in almost the entire spectral range, with detection from the UV (Ortolani \& D'Odorico 1980, B{\"o}hm et al. 1993) and X-rays (Pravdo et al. 2001) to the optical (Solf et al. 1988), infrared (Gredel 1996, Eisl{\"o}ffel et al 2000, Nisini et al. 2005) and radio continuum (Rodriguez et al. 2000). High-angular resolution observations were obtained in the optical with the {\it Hubble Space Telescope} images (e.g. Hartigan et al. 2011, Raga et al. 2015).

We presented the X-shooter observations of HH\,1 in a previous paper (Giannini et al. 2015, hereafter Paper\,I), where we have analyzed some very bright [\feii] lines in order to derive empirically their Einstein $A$-coefficient ratios. In this paper we present the complete UV, optical, and near-infrared spectrum aiming at:
1) providing the most complete atlas of the chemical elements detected in a  protostellar shock; 2) obtaining a complete picture of the gas physical conditions, to be used as a benchmark for shock models in subsequent studies; 3) estimating the chemical abundance of the detected species and inferring on the ability of shocks in destroying dust.

The paper is organized as follows. In Section \ref{sec:sec2} we describe the observations and data reduction and present the HH\,1 X-shooter spectrum. In Section \ref{sec:sec3} we describe the diagnostic tools used to derive the gas physical conditions. These latter, along with the elemental abundances of the detected species, are presented in Section \ref{sec:sec4}. General comments and final remarks are then given in Section \ref{sec:sec5}. In Appendix\,\ref{sec:appendix} we give the complete list of the detected lines.

\section{Observations and data reduction}{\label{sec:sec2}
A summary of the observations of HH\,1 was already given in Paper\,I. For reader's convenience, we repeat here the most important pieces of information. 
We collected 7 X-shooter (Vernet et al. 2011) observations of HH\,1 (pointing: $\alpha_{J2000.0}$= 05$^{h}$ 36$^{m}$ 20{\fs}27, $\delta_{J2000.0}$= $-$06{\arcdeg} 45{\arcmin} 04{\farcs}84), in the period November 1 2013 - January 2 2014, each integrated  30 min on-source. The 11{\arcsec} slit covers the bow-shock head and part of the bow-flanks, as shown in Figure\,\ref{fig:fig1}. To properly subtract the sky emission we acquired 
off-source exposures at 15{\arcsec} from the target, integrated for the same time as on-source. 
The slit, aligned with the direction of the bow-shock (position angle of 129{\arcdeg}), was set to achieve a resolving power of 9\,900,  18\,200 and 7\,780 for the UVB (3\,600 - 5\,900 \AA), VIS (5\,450 - 10\,200 \AA) and NIR arm (9\,900-24\,700 \AA), using slit widths of 0\farcs 5,  0\farcs 4,  and 0\farcs 6, in the three spectrograph arms respectively. The pixel scale is 0\farcs 16 for the UVB and VIS arms, and 0\farcs 21 for the NIR arm. 

The data reduction was accomplished independently for each arm using the X-shooter pipeline v.2.2.0 (Modigliani et al. 2010),  which provides 2-dimensional spectra, both flux and wavelength calibrated. Each of the seven exposures was analyzed separately by means of post-pipeline procedures available within the {\it IRAF} package to subtract the sky-exposure (that presents only telluric lines) and to correct for the relative motions between the Earth and the target at the time of the observation. 
Each NIR image was divided by the spectrum of a telluric standard star taken immediately after the observation, once corrected for both the stellar continuum shape and the intrinsic absorption features (hydrogen recombination lines).  Finally, we combined the seven observations applying a median filter to get the final 2-D image.
To extract the spectrum, we analyzed the spatial profile of the brightest lines, some of which are shown in Figure \ref{fig:fig2}.
The emission of all the lines extend over the entire slit length, with a major emission peak of the atomic lines at a distance of about 2\arcsec - 3\arcsec\, from the bow-shock edge. Differences among the spatial profiles will be commented further on (Section\,\ref{sec:sec4.1}), here we just note that
 most of the emission is found  between 1\farcs5 and  8\farcs8 from the bow-edge, as evidenced  with red lines in Figure\,\ref{fig:fig2}. Since our primary goal is to detect lines from low-abundant species, we integrated the emission over all this zone. The final, 1-dimensional spectrum is shown in Figures\, \ref{fig:fig3}-\ref{fig:fig5}.
This latter lacks a detectable continuum emission (with the exception of a faint emission at the UV wavelengths), and therefore we could not check directly the goodness of the flux inter-calibration between adjacent arms. Rather, we compared fluxes of lines present in the overlapping portions of the spectral segments. While fluxes of lines in both UVB and VIS segments agree within few percent, the flux of the unique bright line in common between the VIS and NIR arm (\hi\,$\lambda$10048\AA\,) differs of 40\,\%. To re-align the spectrum we took as a reference the VIS segment: this potentially introduces an uncertainty in the absolute fluxes, which was taken into account in the few cases when flux ratios of UVB/VIS lines with NIR lines are considered in the analysis.

\subsection{Detected lines}{\label{sec:sec2.1}
In total, we detect more than 500 lines, being the spectrum
particularly rich in [\feii] and \htwo\, lines. In a deeper detail, we detect fine structure ionic lines from atoms with Z $\le$ 28  and with excitation energy up to $\sim$ 40\,000 cm$^{-1}$, \hi\, recombination lines of the Balmer, Paschen, and Brackett series, several \he\, recombination lines (plus one \heii\, line), and \htwo\, ro-vibrational lines with v$_{up} \le$ 9 (excitation energy up to $\sim$ 50\,000 cm$^{-1}$). For comparison, Solf et al. (1988) found around 100 lines in their 3700\,\AA - 10830\,\AA\, spectrum: the sensitivity reached with X-shooter led therefore to increase the number of emission line detections of about a factor of five (although within a larger wavelength range). More important, we detect lines from low-abundance elements, like P, Cl, Ti, and Co, together with high-v$_{up}$ \htwo\, lines, which fall mainly in the optical range. The complete list of the detected lines is reported in Appendix\,\ref{sec:appendix} (Tables \ref{tab:tab5} - \ref{tab:tab7}), where
we give, together with the spectral identification, the energy of the upper and lower level, and the measured flux with the associated uncertainty. Apart from \hi\, lines, the large majority of the other lines is not resolved in velocity, and therefore the full-width at half-maximum (FWHM) of the line profile is typically the instrumental one (i.e. $\sim$ 0.5\,\AA\,, 0.7\,\AA\,, 1.5\,\AA\ in the UVB, VIS and NIR arm, respectively).
The flux was derived by means of a Gaussian fit to the line profile, while the associated uncertainty was estimated by multiplying the spectral noise (rms) at line base times the FWHM. At the adopted spectral resolution, some lines are blended each-other  and were not used in the subsequent analysis. Finally, we note that a few, bright lines present a second, fainter and redder component, whose flux was eventually added to that of the principal component to get the total line flux. Since the detailed study of the shock kinematics and geometry is far from our scientific scopes, the analysis of these individual (possibly resolved) line components is not considered further on in the paper.

\section{Diagnostic tools}{\label{sec:sec3}
The analysis of the HH\,1 spectrum consists of the following steps: 1) estimate of the local extinction; 2) determination of the gas temperature and density; 3) determination of the gas fractional ionization. The tools adopted to derive the above quantities are described in the following sections.

\subsection{Extinction}{\label{sec:sec3.1}
The method adopted to derive the local extinction (A$_V$) is described in Paper I. We just recall here that we estimate A$_V$ using pairs of lines coming from the same upper level, whose difference between the observed and theoretical flux ratios is a function only of the local reddening (see e.g. Gredel 1994). We considered flux ratios between \hi\, lines of Balmer and Paschen series and Paschen and Brackett series, 
along with flux ratios of visible and infrared \htwo\, and atomic fine structure lines. Taking into account all the uncertainties and a total-to-selective
  extinction (R$_V$) between 3 and 5, we get 0.0 mag $\le$ A$_V$ $\le$ 0.8 mag. We adopted A$_V$\,=\,0.4 mag to correct the observed fluxes, by applying the reddening law of Cardelli et al. (1989).
Also, the flux uncertainty due to the A$_V$ estimate (0.4 mag)  was statistically added to the $\Delta$Flux of Tables\,\ref{tab:tab5} - \ref{tab:tab7} and taken into account in the line ratios used to derive the gas physical conditions.

\subsection{The excitation model}{\label{sec:sec3.2}
The physical conditions of each ionic species were derived by comparing the line ratios of its brightest lines with the predictions of an excitation model, which adopts the   
Non Local Thermal Equilibrium (NLTE) approximation for line excitation. The equations of the statistical equilibrium are solved considering as processes for levels population the  electronic collisional excitation/de-excitation and the radiative spontaneous decay. Possible influence on line emissivity due to photo-excitation and stimulated emission are discarded at this step of the analysis, but they are considered  {\it a posteriori} examining line ratios particularly sensitive to such processes (Section\,\ref{sec:sec4.1.1}). The free parameters of the excitation model are the electron temperature $T_{\rm{e}}$  and the electron density $n_{\rm{e}}$, which are derived from the measured flux ratios once corrected for the estimated A$_V$. In order to reveal possible temperature and density gradients inside the shock we estimated $T_{\rm{e}}$ and $n_{\rm{e}}$ for each species separately. In particular, the excitation model was developed for all the ionic species for which collisional and radiative rates are available in the literature (all ions but \nai\, and \coii). Air frequencies and radiative rates are taken from 'The atomic line list'\footnote{v2.04, http://www.pa.uky.edu/~peter/atomic/} and from the 'NIST' database\footnote{http://www.nist.gov/pml/data/asd.cfm}, with the exception of the radiative rates of \crii\, which  were taken from Quinet 1997. 
In Table\, \ref{tab:tab1} we summarize the details of each model. In particular, in the second column we report the average IP$_{ave}$=(IP$_{i-1}$+IP$_i$)/2 between the ionization potentials of the ionic stage {\it i}-1 and {\it i}, to have an idea of the  energy range corresponding to the ionization stage {\it i} of the considered species.  We also give the number of levels considered in the model (column 3), the literature reference for collisional rates with electrons  (column 4) and the temperature range in which these rates have been computed, which, roughly speaking, covers higher temperatures for species with higher IP$_{ave}$ (column 5). In agreement with this occurrence, we have constructed a grid of model solutions in the range: 3000 K $\le T_{\rm{e}} \le$ 20\,000 K ($\delta T_{\rm{e}}$= 1000 K), 5000 K $\le T_{\rm{e}} \le$ 30\,000 K ($\delta T_{\rm{e}}$= 1000 K), and 10\,000 K  $\le T_{\rm{e}} \le$ 80\,000 K
($\delta T_{\rm{e}}$= 5000 K) for ions having IP$_{ave}$ less than 20 eV, between 20 and 30 eV, and higher than 30 eV, respectively. All the grids cover the range 
10$^3$ cm$^{-3} \le n_{\rm{e}} \le$ 10$^7$ cm$^{-3}$ (in steps of log$_{10}$($\delta n_{\rm{e}}$/cm$^{-3}$)\,=\,0.1). We make available these grids, containing the emissivities 
of selected lines, at the website: http://www.oa-roma.inaf.it/irgroup/line$\_$grids/Atomic$\_$line$\_$grids/Home.html.
 
We were able to derive an estimate of $T_{\rm{e}}$ and $n_{\rm{e}}$ from line ratios of most of the observed species 
 with the exceptions of  \clii\,, \pii\, and \tii. Indeed, \clii\, and \pii\ lines come from the same upper level and therefore their ratio is independent from the gas
  physical conditions,  while lines of \tii\, have a too low signal-to-noise ratio to provide meaningful determinations. Models for these species were however constructed to compute the intrinsic line emissivities once assumed the physical parameters, since these predictions are needed to derive the chemical abundances (see Section\,\ref{sec:sec4.4}).

\subsection{The ionization model}{\label{sec:sec3.3}
To derive the fractional ionization $x_{\rm{e}}$ = $n_{\rm{e}}$/$n_{\rm{H}}$ (where $n_{\rm{H}}$=$n_{\rm{H^0}}$+$n_{\rm{H^+}}$) and the relative abundances X$^i$/X$^{i+1}$ of the atomic species in different ionization stages, we developed a ionization code that takes into account the following
processes: collisional ionization, radiative and dielectronic recombination, and direct and inverse charge-exchange with Hydrogen. Ionization, radiative and dielectronic recombination rates are taken from Landini \& Monsignori-Fossi (1990) with the exception of the Iron data (Arnaud \& Raymond 1992).  Charge-exchange data are from Kingdon \& Ferland (1996) apart from Carbon data (Stancil et al. 1998). Dielectronic recombination coefficients at low temperatures have been considered in O, N, and C models (Nussbaumer \& Storey 1983). Notably, while the ionization and recombination processes are function only of the electron 
temperature, direct and inverse charge-exchange  depend also on the fractional ionization $x_{\rm{e}}$. Out of the species detected in different ionization stages, charge-exchange rates are relevant for the processes N$^{0}$ + H$^{+}$ $\rightleftarrows$  N$^{+1}$ + H$^{0}$,  Fe$^{+1}$ + H$^{+}$ $\rightleftarrows$  Fe$^{+2}$ + H$^{0}$, O$^{0}$ + H$^{+}$ $\rightleftarrows$  O$^{+1}$ + H$^{0}$, and O$^{+1}$ + H$^{+}$ $\rightleftarrows$  O$^{+2}$ + H$^{0}$.

Since different ionic stages are found in different regimes of temperature and density (Section\,\ref{sec:sec4.1}), we have fitted $x_{\rm{e}}$ (0 $\le x_{\rm{e}} \le$ 1 in steps $\delta x_{\rm{e}}$ = 0.05) in the range of $T_{\rm{e}}$ and $n_{\rm{e}}$  in which adjacent ionization stages are expected to co-exist in significant percentages. The fit results and their implications are discussed in Section\,\ref{sec:sec4.2}.

\section{Results and Discussion}{\label{sec:sec4}

\subsection{Density and temperature stratification}{\label{sec:sec4.1}
Most of the observed lines come from the first 5 fine structure levels of the ground state. For these species 5-level models have been considered, and  
$T_{\rm{e}}$ and $n_{\rm{e}}$ have been derived from diagnostic diagrams involving the brightest lines. Examples are given in Figure\,\ref{fig:fig6}. For complex atomic systems, e.g. \feii, \feiii, and \NIii, we have constructed more complicated models, and derived the physical parameters following the best-fitting method proposed by Hartigan \& Morse (2007), which consists in iteratively changing the line used for the normalization to minimize the $\chi^2$ value of the fit. The best fits are shown in Figure\,\ref{fig:fig7}.

Temperature and density fitted by the excitation model in each ion are reported in Table\,\ref{tab:tab1} (columns 6 and 7). 
For the large majority of the ions, the derived $T_{\rm{e}}$ falls inside the range of validity of the model (column 5). A remarkable exception is \oii\, whose  temperature was derived by extrapolating the collisional rates from their maximum value. With reference to  Figure\,\ref{fig:fig6} we note that while for some ions we are able to precisely derive  both $T_{\rm{e}}$ and $n_{\rm{e}}$ (e.g. \n, \oii, \sii), for others we can significantly constrain only one of the two parameters (e.g. \ci, \oiii). However, for the highly ionized species, the information derived from the line ratios can be integrated with others constraints. This is, for example, the case of \ariv\, shown in the bottom-left panel of Figure\, \ref{fig:fig6}. Indeed, while the ratio 4740\AA /4711\AA\, is a good indicator of the density, it is almost independent from the local temperature. However, this latter can be evaluated adding as a constraint the Argon abundance X(Ar) derived from the [\ariii] lines (X(Ar)\,=\,1.1 10$^{-5}$ $\div$ 1.7 10$^{-5}$ X(H), see Section\,\ref{sec:sec4.4}).  Taking this range, we get that the observed fluxes of the \ariv\, lines are fitted by the excitation model only if $T_{\rm{e}}$(\ariv) is in the range 75\,000 K $\div$ 85\,000 K. For example, if $T_{\rm{e}}$(\ariv) were $\la$ 60\,000 K, X(Ar) should exceed the solar value by an order of magnitude. Similarly, we find that to reproduce the observed lines of \neiii\, at temperatures lower than $\sim$ 50\,000 K, X(Ne) should  exceed the solar abundance by about a factor of 5. More reasonably, imposing that  X(Ne) $\la$ X(Ne)$_\odot$, we put a lower limit of  60\,000 K to the \neiii\, temperature.
 
From Table\,\ref{tab:tab1} it can be noted that higher $T_{\rm{e}}$ are found for ions with higher  IP$_{ave}$. This is better shown in Figure\, \ref{fig:fig8}, where we plot IP$_{ave}$ versus $T_{\rm{e}}$ finding  a significant correlation (regression coefficient $R$=0.95) between the two quantities. The linear best fit through the data points is Log($T_{\rm{e}}$) = 0.02  IP$_{ave}$ + 3.8. Hints of a correlation were already found by Solf et al. (1988, their
figure 2), although with a temperature plateau at $\sim$ 35\,000 K above 30 eV where no more data points were available. With the new X-shooter observations we extend the correlation up to $\sim$ 80\,000 K, i.e. at temperatures expected in the recombination region immediately behind a fully dissociative shock front (J-ump shocks, Hollenbach \& McKee 1989).

The electron density is well constrained by a few  line ratios. Roughly speaking, we can identify
two density regimes: a more tenuous gas component with $n_{\rm{e}}$ between 1.0 10$^3$ cm$^{-3}$ and 5.0 10$^3$ cm$^{-3}$, probed 
by  [\n], [\feii], [\sii], and [\oii] line ratios, and a denser gas component, with $n_{\rm{e}}$ $\ga$ 1.0 10$^5$ cm$^{-3}$, probed by
[\feiii], [\ariv] and possibly [\siii], [\ariii] and [\oiii] line ratios. Therefore, the densest component seems to be associated with the highest temperatures. With reference to Figure\,\ref{fig:fig2}, this scenario is consistent with the presence of two partially mixed ionized gas components, one spatially narrower with $T_{\rm{e}}$  $\ga$ 30\,000 K and $n_{\rm{e}}$  $\ga$ 1.0 10$^5$ cm$^{-3}$ located close to the shock front (roughly peaking at 2\arcsec\,- 3\arcsec\, from the bow-shock edge), and a second, rear and broader component with lower temperature and electron density.
We also identify a rather neutral component (namely that emitting the bulk of the \oi\, and \ci\, emission, see Section\,\ref{sec:sec4.2}), which encompasses the entire slit length at a roughly constant level of emission, as shown for example by the [\ci]9850 spatial profile of Figure\,\ref{fig:fig2}. For this component $n_{\rm{e}}$ is not representative of the total gas density, being less than 10\% of the neutral Hydrogen density $n_{\rm{H}}$ (we estimate $x_{\rm{e}}$(\oi) $\sim$ 0). Since the lower limit to $n_{\rm{e}}$ is 10$^3$ cm$^{-3}$ for both \oi\, and \ci\, (Table\,\ref{tab:tab1}), we derive $n_{\rm{H}}$ $>$ 10$^4$ cm$^{-3}$ for this neutral region.

\subsubsection{Role of fluorescence}{\label{sec:sec4.1.1}

In Section\,\ref{sec:sec3.2} the observed line ratios have been analyzed under the assumption that line excitation is due only to collisional processes. Here we explore whether a further contribution from fluorescence excitation may be significant, also considering that a continuum emission, although faint, is present in the very blue part of the HH\,1 spectrum. As shown by Lucy (1995), Bautista et al. (1996) and Bautista \& Pradhan (1998) a powerful way to evaluate the relevance of photo-excitation is to investigate a number of line ratios sensitive to the UV-pumping. The above papers individuate several [\NIii] and
[\feii] flux ratios as particularly suited to this kind of analysis. In Table\,\ref{tab:tab2} the observed, de-reddened ratios (column 2), are compared  with the theoretical predictions for purely collisional excitation (column 3) and for both collisional and fluorescent excitation (column 4). The theoretical values are reported for $T_{\rm{e}}$ = 10\,000 K and for 10$^3$ cm$^{-3}$ $\le$ $n_{\rm{e}}$ $\le$ 10$^5$ cm$^{-3}$. The very good agreement between the observations and the ratios expected for purely collisional excitation led us to conclude that fluorescence has (if any) a very marginal role in line excitation.
As we will show in Section\,\ref{sec:sec4.3}, this result is also confirmed by the analysis of the \htwo\, emission.

Similarly, radiative processes are not at the origin of the faint continuum  observed roughly between 3\,500 \AA\, and 6\,000 \AA\,. Indeed, the fit through the blue part of
the spectrum gives  F$_\lambda \varpropto \lambda^{-n}$, with $n \approxeq$ 3, as expected for two-photon emission processes caused by collisions of hydrogen-like and helium-like ions with free electrons (Dopita et al. 1982; Brugel et al. 1981). Notably, two-photon continuum processes are expected to become important if $n_{\rm{e}}$ $<$ 7 10$^{3}$ cm$^{-3}$ (Mewe et al. 1986), as probed by a number of atomic species (Table\,\ref{tab:tab1}).

\subsection{Ionization fraction}{\label{sec:sec4.2}
The results of the ionization model are presented in Figure \ref{fig:fig9}, where we plot the flux ratios of lines from adjacent ionization stages as a function of $x_{\rm{e}}$
for the three atoms (N, O, Fe) that are sensitive to charge-exchange processes. The top-left panel shows the [\n]5198\,\AA\,/[\nii]6585\,\AA\, ratio vs.  $x_{\rm{e}}$ assuming the  ranges of $T_{\rm{e}}$ and  $n_{\rm{e}}$ derived from the \nii\, excitation model. 
We fit 0.55 $\le$ $x_{\rm{e}}$ $\le$ 0.85. However, since part of the  \n\, emission comes from a region with $T_{\rm{e}}$ $<$ 10$^4$ K (see Table\,\ref{tab:tab1}), only a fraction of the observed [\n]5198 \AA\, line can be attributed to the same gas emitting the [\nii]6585 \AA\, line. In this sense, the observed ratio  has to be regarded as an upper limit, with a consequent shift of $x_{\rm{e}}$ towards higher values.
In the two bottom panels of Figure\,\ref{fig:fig9} we show the results of the  Fe$^{+1}$-Fe$^{+2}$ equilibrium code, which has been solved by assuming the density range derived from [\feiii] lines (note that the dependence  between fractional ionization and electron density is very shallow). Also, since the temperature is practically not constrained by the [\feiii] lines, we fixed $T_{\rm{e}}$ at 10\,000 K. Several flux ratios have been examined, obtaining 0.50 $\le$ $x_{\rm{e}}$ $\le$ 0.70. For the same reasons as those explained for the N$^{0}$-N$^{+1}$ equilibrium, if \feiii\, mostly came from a gas component with temperature higher than 10$^4$ K, all the [\feiii] line fluxes should be considered as upper limits and  $x_{\rm{e}}$  would slightly decrease.
Finally, as far as Oxygen is concerned, we note that the temperature fitted by [\oi] and [\oii] lines are substantially different (see Table \ref{tab:tab1}). 
The ionization code indicates that at $T_{\rm{e}}$ $\sim$ 10$^4$ K  (fitted from \oi\, lines) the percentage O$^0$:O$^{+1}$ is 99:1, therefore, it is likely that the bulk of the [\oi] emission comes from a substantially neutral region. Conversely, the lower limit of 2.0 10$^4$ K derived for the region where Oxygen is singly ionized is compatible with the range of  $T_{\rm{e}}$ derived from [\oiii] line ratios. The ionization equilibrium between O$^{+1}$ and  O$^{+2}$ was therefore solved for the physical conditions got from the [\oiii] lines. As shown in Figure \ref{fig:fig9}, top-right panels, the gas component emitting \oiii\, lines is practically fully ionized.

In Figure\,\ref{fig:fig8} the red points represent $x_{\rm{e}}$ vs. IP$_{i}$, while in Table\,\ref{tab:tab3} we summarize the results of the ionization equilibrium model. 
The first column lists the element for which the code was solved, while the second column contains the ionization stages taken into consideration. In the third column
we report the values of $x_{\rm{e}}$ derived from the above analysis. For the atomic species for which we could not directly derive $x_{\rm{e}}$, we assume the value estimated from the element (among Nitrogen, Oxygen and Iron) with IP$_{ave}$ similar to that of the element under consideration. 

The ionization equilibrium code allows one also to derive the relative abundance of adjacent ionization stages, as a function of $x_{\rm{e}}$ and $T_{\rm{e}}$. We give these quantities in the last column of Table\,\ref{tab:tab3}, computed at the temperatures given in the fourth column (that roughly coincide with the average $T_{\rm{e}}$ derived from the excitation model) and for the $x_{\rm{e}}$ values of the third column. When a range of $x_{\rm{e}}$ is given, X$^i$/X$^{i+1}$ is computed accordingly.
In particular, we note that the equilibrium ionization of He indicates that a relevant percentage of He$^{+1}$ is found only if the gas temperature exceeds $\sim$ 30\,000 K (in agreement with the relation of Figure\,\ref{fig:fig5}, being IP$_{ave}$(He)\,=\,39.5 eV). This result, as well as the relative abundances, will be used in the following to derive the atomic abundances (Section\,\ref{sec:sec4.4}).

\subsection{Physical conditions of the molecular component}{\label{sec:sec4.3}
As anticipated in Section\,\ref{sec:sec2.1}, the spectrum of HH\,1 is extremely rich in \htwo\, line emission, observed both in the VIS and NIR segments. In total, around 200 ro-vibrational lines with  1 $\le$ v$_{up}$ $\le$ 9 are detected. In total, around 200 ro-vibrational lines with  1 $\le$ v$_{up}$ $\le$ 9 are detected. Line identifications and fluxes are reported in Table\,\ref{tab:tab7}. The \htwo\, emission was analyzed in the framework of a Boltzmann
diagram, i.e. a plot of ${\rm ln}~(N_{(v,J)}/g_Jg_s)$ against E$_{(v,J)}/k$, where N$_{(v,J)}$ (cm$^{-2}$) is the column density of level (v,J), E$_{(v,J)}$/k (K) its excitation energy and g$_J$g$_s$= (2J+1)(2I+1) its statistical weight. In the assumption of optically thin emission, ortho-to-para (o/p) ratio of 3,  and thermalization at a single temperature, the points in the diagram fall onto a straight line, while for a range of kinetic temperatures the Boltzmann diagram forms a curve. Moreover, data points coming from the same upper level are expected to superpose in the diagram once de-reddened. The Boltzmann diagram of the \htwo\, lines detected with a snr $\ge$ 5 is represented in Figure\,\ref{fig:fig10}, top panel. To minimize the uncertainties we have also discarded lines close in wavelengths to the atmospheric holes or affected by artifacts (e.g. sky residuals or bad pixels, mainly present in the NIR part of the spectrum). The best fit  was obtained optimizing at the same time both the A$_V$ value and the gas temperature. We get A$_V$=0.5 mag, which has been applied to all lines.
The temperature is found as the reciprocal of the slope of the straight line through the data points. If all the lines are considered (solid line) we get an average temperature over the whole H$_2$ region of $\sim$ 5\,500 K. However, columns of low-excited lines (typically lines with $v_{up}$=1 and E$_{(v,J)}$ $\le$ 15\,000 K) are severely underestimated by this fit. Rather, the overall emission is better reproduced considering two temperature components, at  $\sim$ 2\,500 K and $\sim$ 6\,300 K (dashed lines). We note that the first temperature component is in quite good agreement with the determination of 2750 K obtained by Gredel (1996) on the base of the NIR spectrum only.

The analysis of the \htwo\, emission suggests some considerations. First, the fact that the data points are fitted under LTE conditions, indicates that the gas density must be higher than the critical densities of the levels from which the lines originate (to have an indication, $n_{\rm{crit}}$ $\ga$ 10$^4$ cm$^{-3}$ at T = 3\,000 K for most of the levels). Therefore, as derived for the neutral gas component, also the molecular emission should arise from a region with density exceeding 10$^4$ cm$^{-3}$.

Second, we note that a the fitted temperatures the \htwo\, gas is expected to have reached equilibrium between the ortho and para form, corresponding to  o/p = 3. 
We have verified this circumstance, since in the opposite case the data points on the Boltzmann diagram would anymore align onto a straight line, but rather would follow a well defined 
zig-zag distribution (e.g. Nisini et al. 2010).

Third, both the high temperature fitted and the detection of lines with high excitation energy (up to  E$_{(v,J)}/k$ $\sim$ 50\,000 K) may indicate that the \htwo\, emission is at least partially influenced by ultraviolet pumping. Indeed, the strong ultraviolet continuous emission detected in the range 1\,400 \AA\, $\le$ $\lambda$ 1\,600 \AA\, by B{\"o}hm et al. (1993) was likely attributed to a fluorescent \htwo\, continuum.
 We compared our data with the model by Sternberg \& Dalgarno (1989) who have predicted the fluxes of bright \htwo\, lines for increasing electron density and for a local radiation field a factor of one hundred higher than the interstellar field ($\chi$ parameter). We plot in Figure\,\ref{fig:fig10}, second to bottom panel, the ratios of the observed lines over their theoretical values for  10$^3$ cm$^{-3}$ $\le n \le$ 10$^6$ cm$^{-3}$, being $n=n_{\rm{H_2}}$ + $n_{\rm{H}}$. A marginal agreement with the theoretical predictions is found only for high-density values, at which, however, collisional excitation is expected to have a major role. Therefore, we can conclude that fluorescence is fairly negligible  in \htwo\, line excitation (at least as far as the observed wavelength range is concerned), thus confirming what found also for the atomic emission in Section\,\ref{sec:sec4.1.1}.

Finally, if we consider the gas as purely collisionally excited, we can derive some hints on the nature of the shock from the rotational diagram, giving the fact that 
the level populations are sensitive to the shock parameters  (pre-shock density, shock velocity and age, and strength of the magnetic field). In particular, 
the presence of a magnetic field acts to dampen and broaden the shock wave; therefore, an increased magnetic field strength enhances the departure from LTE (e.g. Flower et al. 2003). Conversely, a J-type shock is associated to a narrow shock wave, more easily thermalized and with higher column densities of the levels with larger vibrational numbers. In a very qualitative way, this second scenario appears therefore to be more suitable in representing the observed \htwo\, emission.

\subsection{Chemical abundances}{\label{sec:sec4.4}
Chemical abundances of the atomic species have been obtained by comparing fluxes of bright lines with the flux of the H$\beta$ line. Indeed, the atomic abundance is a direct function of the observed flux ratio according to the relationship: 

\begin{eqnarray}
\frac{X}{H} = \frac{F_{line}}{F_{H\beta}}\times\left[\frac{\epsilon_{line}(T_{\rm{e}},n_{\rm{e}})}{\epsilon_{H\beta}((T_{\rm{e}},n_{\rm{e}})} \frac{X^i}{X} \frac{H}{H^+}\right]^{-1} 
\end{eqnarray}

where $X/H$ is the
abundance of the $X$ atomic species relative to Hydrogen, $F_{line}/F_{H\beta}$ is the observed (de-reddened) flux ratio between a line of the X$^i$ ion and H$\beta$, $\epsilon_{line}(T_{\rm{e}},n_{\rm{e}})$
and $\epsilon_{H\beta}(T_{\rm{e}},n_{\rm{e}})$ are the line and H$\beta$ emissivities computed at the physical conditions derived in Table\,\ref{tab:tab1}, ${X^i}/{X}$ is the percentage of the atom $X$ in the $X^i$ ionization stage, and  $H/H^+$ is the reciprocal of $x_{\rm{e}}$.\\
We selected the brightest lines observed in each species and applied the above formula to derive  $X/H$. While  $\epsilon_{line}(T_{\rm{e}},n_{\rm{e}})$ is an output of our excitation
model, $\epsilon_{H\beta}(T_{\rm{e}},n_{\rm{e}})$ is tabulated by Storey \& Hummer (1995) under case B approximation, whose validity has been demonstrated in Paper I. As shown in the same paper, the bulk of Hydrogen line emission comes from gas at temperatures not exceeding $\sim$ 20\,000-25\,000 K, so that the above formula cannot be applied to the highest ionized
atoms (i.e. S$^{+2}$, Ar$^{+2}$, O$^{+2}$) which are excited at much higher temperatures. To compute the abundance of these species we have therefore taken as a  
reference the \heii\, 4-3 line, which is expected to be excited at $T_{\rm{e}}$ $\ge$ 30\,000 - 40\,000 K (see Table\, \ref{tab:tab3}). and whose emissivity as a function of $T_{\rm{e}}$ and $n_{\rm{e}}$ is taken from Storey \& Hummer (1995). 
Abundance with respect to Hydrogen is then derived by assuming  X(He)/X(H)=0.0995, measured by Esteban et al. (2004) in the Orion nebula. Similarly, since Magnesium and Carbon emit mainly in neutral and low-temperature zones inside the shock, we compute X(Mg) and X(C) taking as a reference the [\oi]6300\AA\, line, assuming the Oxygen abundance derived from [\oiii] lines and considering that also the [\oi] lines come from a region where $x_{\rm{e}}$ $\sim$ 0, as shown in Section\,\ref{sec:sec4.2}.

The derived chemical abundances are given in Table\,\ref{tab:tab4}, column 2. We give for each species a range of values, which are obtained by computing the line emissivities for the extreme 
physical conditions of Table\,\ref{tab:tab1}. The abundances of Phosphorus and Chlorine, for which we were unable to directly derive $T_{\rm{e}}$ and $n_{\rm{e}}$, were estimated by assuming the conditions of ions with similar  IP$_{ave}$, i.e. Fe$^{+1}$ and S$^{+1}$, respectively.

In columns  3-4 we compare the abundances derived in HH\,1  with the solar values. First, we note that most of our determinations are below the solar abundances. This is indeed a not surprising result, since abundances below the solar ones are found in different environments of our Galaxy (like diffuse clouds,
 \hii\, regions, surroundings of B-type stars, see Wilson \& Rood 1994 and Savage \& Sembach 1996), and, in particular, in the atmospheres of young stars in the Gould's Belt (Spina et al. 2014, Biazzo et al. 2012).

 Non-refractory species are, on average, $\sim$ 0.15 dex under-abundant, while species commonly bound on dust grains present an average deviation of $\sim$ 0.5 dex, being Magnesium and Calcium the most depleted species. Such result proves the presence of dust inside the shock.
Several studies have been done to evaluate the degree of depletion of the refractory species. In particular, sparse Iron 
 depletion factors have been estimated in shock regions, going from around 80\% (Mouri \& Taniguchi 2000. Podio et al. 2006), to intermediate values between 40\% and 70 \% (Nisini et al. 2002, Giannini et al. 2008, 2013) up to negligible percentages (B{\"o}hm \& Matt 2001, Pesenti et al. 2003). In this respect, the depletion factor of $\sim$ 40\% found in HH\,1, represents an intermediate case.

An interesting comparison of the found abundances is in principle with the determinations obtained inside the Orion nebula, since the possible differences with this latter would directly reflect the effect of the shock passage through the surrounding medium. However, the differences we find with the determinations by Esteban et al. (2004. columns 5-6), mainly for C, Ar and O, are likely more due to a different computational approach than to intrinsic abundance fluctuations. For example, X(C) was computed by Esteban et al. by means of recombination lines, which, as shown by the same authors, always give larger estimates than the forbidden lines.  Also, the high, over-solar,  X(Ar) measured in that paper, likely arises to the low temperature, not exceeding $\sim$ 10\,000 K, estimated for the emitting gas. The only species for which the abundance difference between HH\,1 and the Orion nebula is really significant is represented by X(Fe). In the Esteban et al. work, X(Fe) is computed under different hypotheses on Iron fractional ionization, but in any case X(Fe) is a factor $\sim$ 5$\div$ 25 lower than in HH\,1. Such a major difference likely indicates that Iron is highly depleted  inside the nebula while partially released in gas-phase inside the shock. 
 
More in detail, our result can be compared with specific studies done on HH\,1 in previous works. Beck-Winchatz et al. (1994, 1996) have computed the Iron depletion degree in HH\,1 by comparing [\feii] with [\oi] and [\sii] line fluxes,
finding abundance ratios similar to the solar ones. As in the case of the Orion nebula, the partial discrepancy with the present result could be due to the approximation on the physical conditions (one single temperature and density) and fractional ionization assumed in those works. 

Finally, it is interesting to compare the abundances in the HH\,1 bow-shock with those determined in the jet for the refractory species by Nisini et al. (2005) and Garcia Lopez et al. (2005).
Hints of a selective depletion were evidenced in those papers by comparing the ratios of different species in knots located at increasing distances from the jet driving source. For example, assuming for all the species the solar abundance, the Iron depletion degree decreases from 80\% to $\sim$ 30\% from the internal to the external knots, while  
the ratio [\feii]16435/[\tii]21599 increases from 150 to 280, implying a gas-phase X(Fe)/X(Ti) ratio from  3.3 to 1.8 times lower than the solar value (corresponding to [\feii]16435/[\tii]21599  $\simeq$ 500). Notably, in the bow-shock the Iron depletion factor is roughly the same as the external knots, but the above line ratio is $\simeq$ 500. In a very simple view, these observational evidences can only be explained 
with an 'over-abundance' of the gas-phase Titanium with respect to Iron inside the jet, while in the bow-shock the two species are depleted in similar percentages.  

\section{Final remarks}{\label{sec:sec5}
We have presented the 3\,600 -24\,700 \AA\, deep X-shooter spectrum of the HH\,1 bow-shock, one of the brightest known Herbig-Haro objects. Our main aims are to provide an atlas of the
chemical species involved in a protostellar shock along with a detailed picture of the gas physical conditions to be used as a benchmark for shock models. Our analysis and main results 
can be summarized as follows:
\begin{itemize}
\item[1.] The spectrum is extremely rich in line emission, with more than 500 detected lines coming from atomic species with Z up to 28, Hydrogen and Helium recombination lines, and
\htwo\, ro-vibrational lines with v$_{up}$ $\le$ 9. 
\item[2.] The atomic emission was analyzed with an excitation code under NLTE approximation. We are able to precisely determine both the temperature and density of the emitting gas for a large number of species, obtaining a detailed view of the regions where the lines of the ion in question arise. We find a well defined correlation between the electron temperature and the increasing ionization energy (IP$_{ave}$), going from less than 10$^4$ K up to $\sim$ 8 10$^4$ K  for 4 eV $\la$ IP$_{ave}$ $\la$ 60 eV. In this range we fit Log $T_{\rm{e}}$ = 0.02 IP$_{ave}$ + 3.8.
The high temperatures found are expected by theoretical models of fully dissociative 
shocks, which predict the lines to form in the recombination region behind the shock front. In this view, collisions are at the origin of line formation, as we have proven also examining
flux ratios of lines particularly sensitive to UV-pumping. While the temperature progressively increases with the ionization energy, two distinct density regimes are individuated for the ionized gas:  a more tenuous and spatially broader component ($n_{\rm{e}}$ $\sim$ 10$^3$ cm$^{-3}$) roughly associated with the gas at $T_{\rm{e}}$ $\la$ 30\,000 K, and a denser, compact component ($n_{\rm{e}}$ $\ga$ 10$^5$ cm$^{-3}$) associated with the gas at $T_{\rm{e}}$ $\ga$ 30\,000 K, likely located in the compression zone close to the edge of the shock front. A further neutral component is also evidenced, at temperature $\la$ 10\,000 K and density $>$ 10$^4$ cm$^{-3}$. 
\item[3.] H$_2$ lines were analyzed by means of the Boltzmann diagram, which reveals the presence of two different molecular components, fairly thermalized at temperatures  $\sim$ 2\,500 K and $\sim$ 6\,300 K. The brightest lines were compared with the expectations of a fluorescence model, substantially finding inconsistence with this excitation mechanism. Rather, a qualitative analysis suggests the H$_2$ emission, as the atomic one, to arise in the molecular recombination zone of a (at least partially) dissociative shock.
\item[4.] The gas fractional ionization was measured by solving the ionization equilibrium equations for adjacent ionization stages of each atomic species. As found for the temperature, also the fractional ionization 
rapidly increases with the ionization energy. In particular, for IP$_{ave}$ $\sim$ 30 eV, the gas is fully ionized.
\item[5.] Chemical abundances of C, N, O, Mg, P, S, Cl, Ar, Ca, Ti, Cr, Fe, and Ni have been derived from ratios of bright lines with respect to the H$\beta$ and \heii\, 4-3 line. With respect to the Sun, non-refractory species result on average 0.15  dex under-abundant. Refractory species are even less abundant (average factor of 0.5 dex) likely because they are not all in gas-phase but rather locked on the mantles of dust grains present inside the shock. In particular, the Iron depletion factor ($\sim$ 40\%) is intermediate with respect to the estimates obtained in a number of proto-stellar environments. The effect of a selective depletion, already put into evidence by previous studies of the HH\,1 jet, is confirmed  in the bow-shock. According to this effect, Iron is progressively less depleted going from the inner parts of the jet towards increasing distance from the driving source, while Titanium has a more constant (if not opposite) behavior. 
\end{itemize}


\acknowledgments
The ESO staff is acknowledged for support with the observations and the X-shooter pipeline.  The authors thank J. M. Alcal\'a for his useful comments and suggestions.
SA acknowledges support from the T-REX-project, the INAF (Istituto Nazionale di Astrofisica) national project aimed at maximizing the participation of astrophysicists and Italian industries 
to the realization of the E-ELT (European Extremely Large Telescope). The T-REX project has been approved and funded by the Italian Ministry for Research and University (MIUR) in the framework of  the “Progetti Premiali 2011”  and then “Progetti Premiali 2012”
Financial support from INAF, under  PRIN2013 program 'Disks, jets and the dawn of planets' is also acknowledged.
LP has received funding from the European Union Seventh Framework Programme (FP7/2007-2013) under grant agreement n. 267251.



{\it Facilities:} \facility{VLT (X-shooter)}}.

\clearpage



\begin{figure}
\plotone{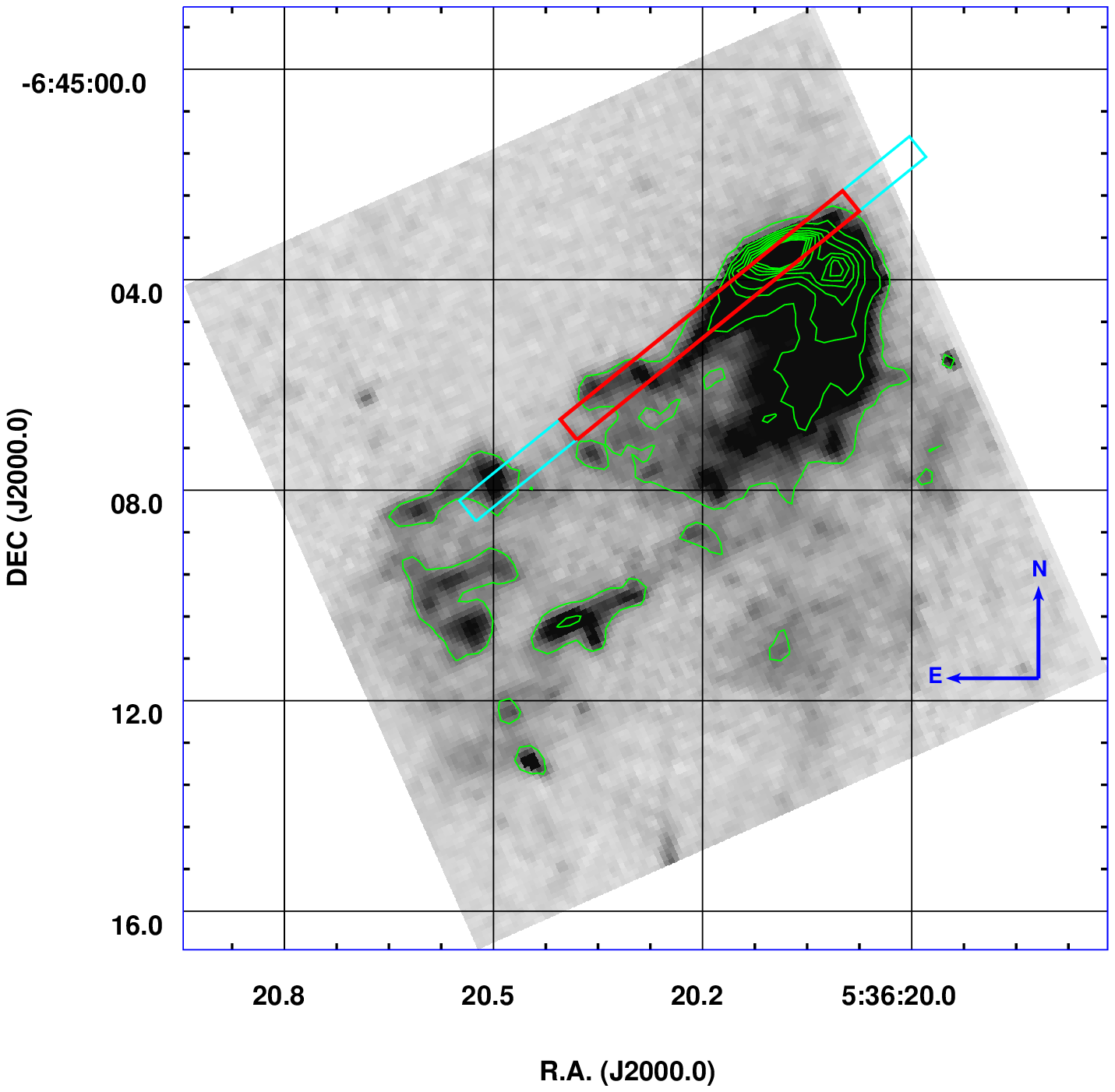}
\caption{HH\,1 seen in H$\alpha$ with the Hubble Space Telescope (image taken from the public archive). Contour levels  go from 90\,\%  to 100\,\% of the intensity peak (not flux calibrated). Cyan box represents the X-shooter slit while red box indicates the spatial region where the spectrum has been extracted. \label{fig:fig1}}
\end{figure}

\begin{figure}
\epsscale{.80}
\plotone{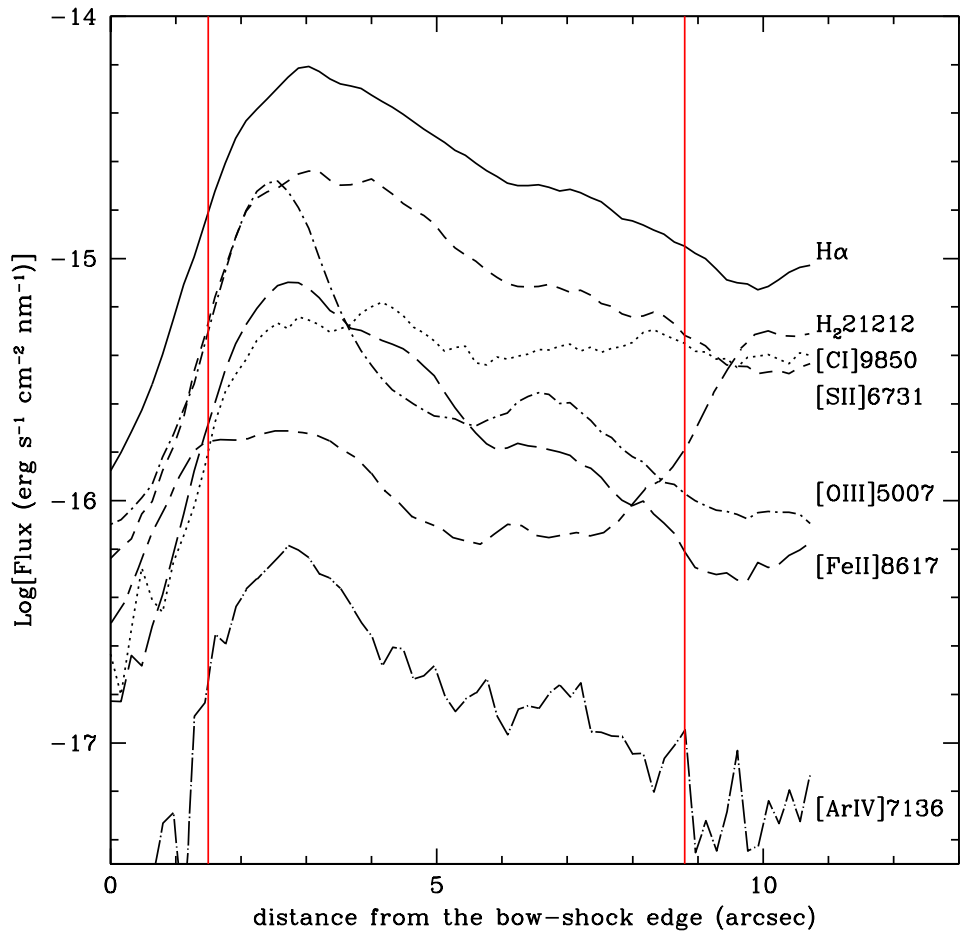}
\caption{Examples of spatial profiles of bright lines along the HH\,1 region covered by the X-shooter slit. The origin of the x-axis is the bow-shock edge. The selected lines are: H$\alpha$ (solid); \htwo21212 (short-dashed long-dashed); [\ci]9850 (dotted); [\sii]6731 (short-dashed);  [\oiii]5007 ((dotted short-dashed)); [\feii]8617 (long-dashed);  [\ariv]7136 (dotted long-dashed). The red vertical lines delimit the zone where the spectrum has been extracted.\label{fig:fig2}}
\end{figure}

\begin{figure}
\epsscale{1.02}
\plotone{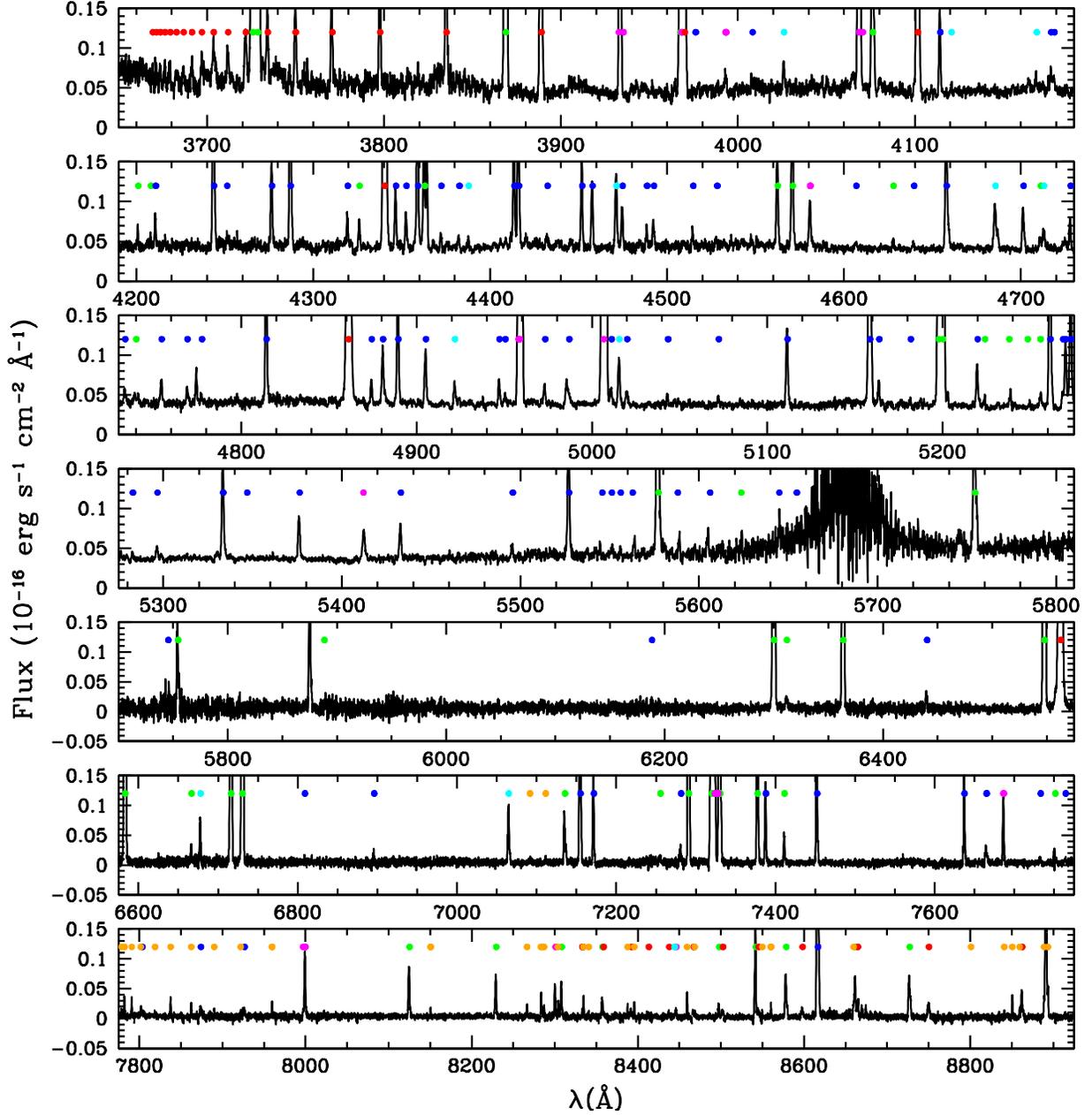}
\caption{X-shooter spectrum of HH\,1 in the range 3650$-$8925 \AA. Lines from different species are marked with different colors: green: atomic lines (apart from Iron lines); blue: Iron lines; red: Hydrogen recombination lines, cyan: Helium recombination lines; orange: \htwo\, ro-vibrational lines, magenta: blends. \label{fig:fig3}}
\end{figure}

\begin{figure}
\epsscale{1.02}
\plotone{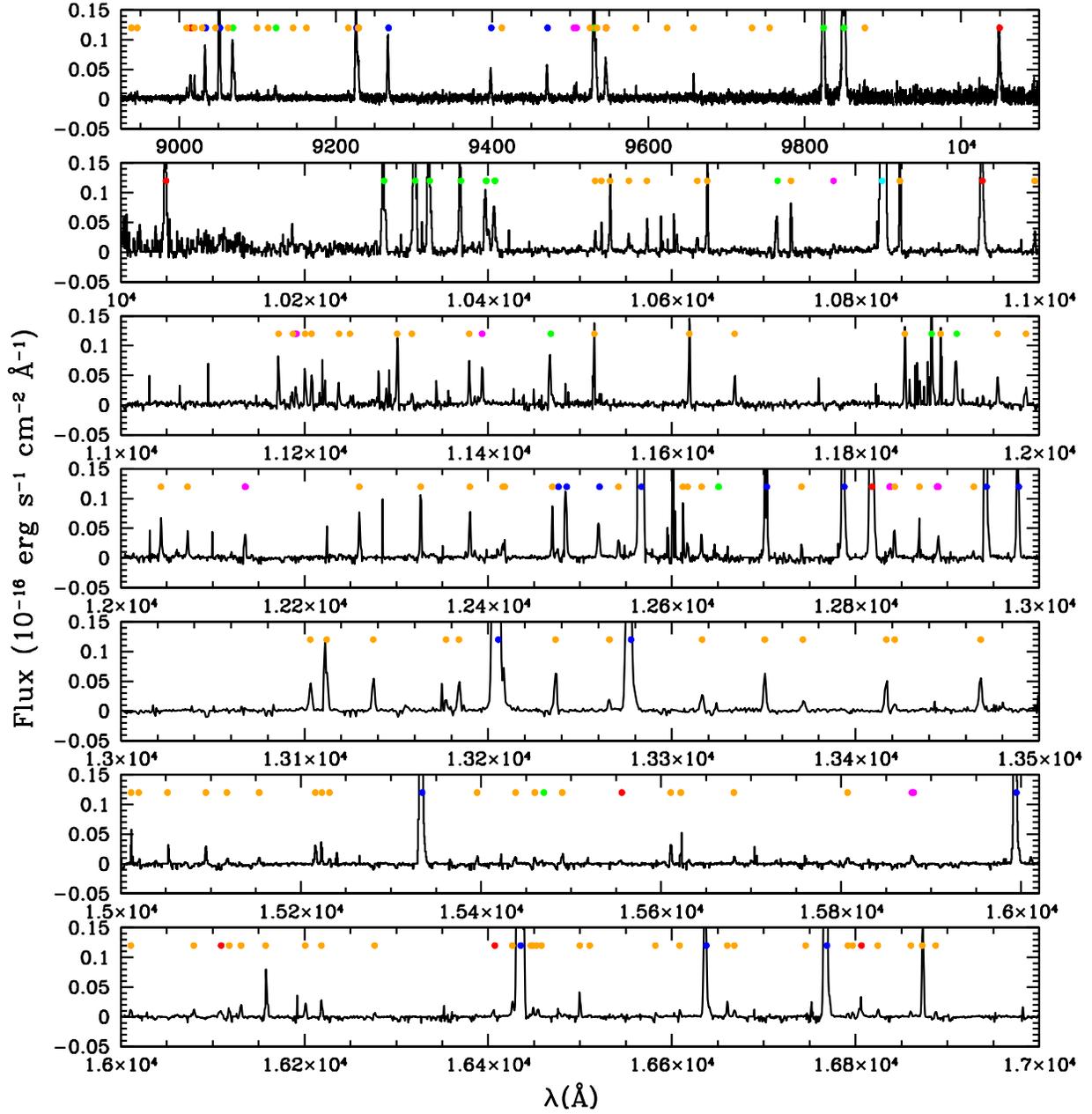}
\caption{As in Figure\,\ref{fig:fig3} in the range  8925$-$17000 \AA. \label{fig:fig4}}
\end{figure}

\begin{figure}
\epsscale{1.02}
\plotone{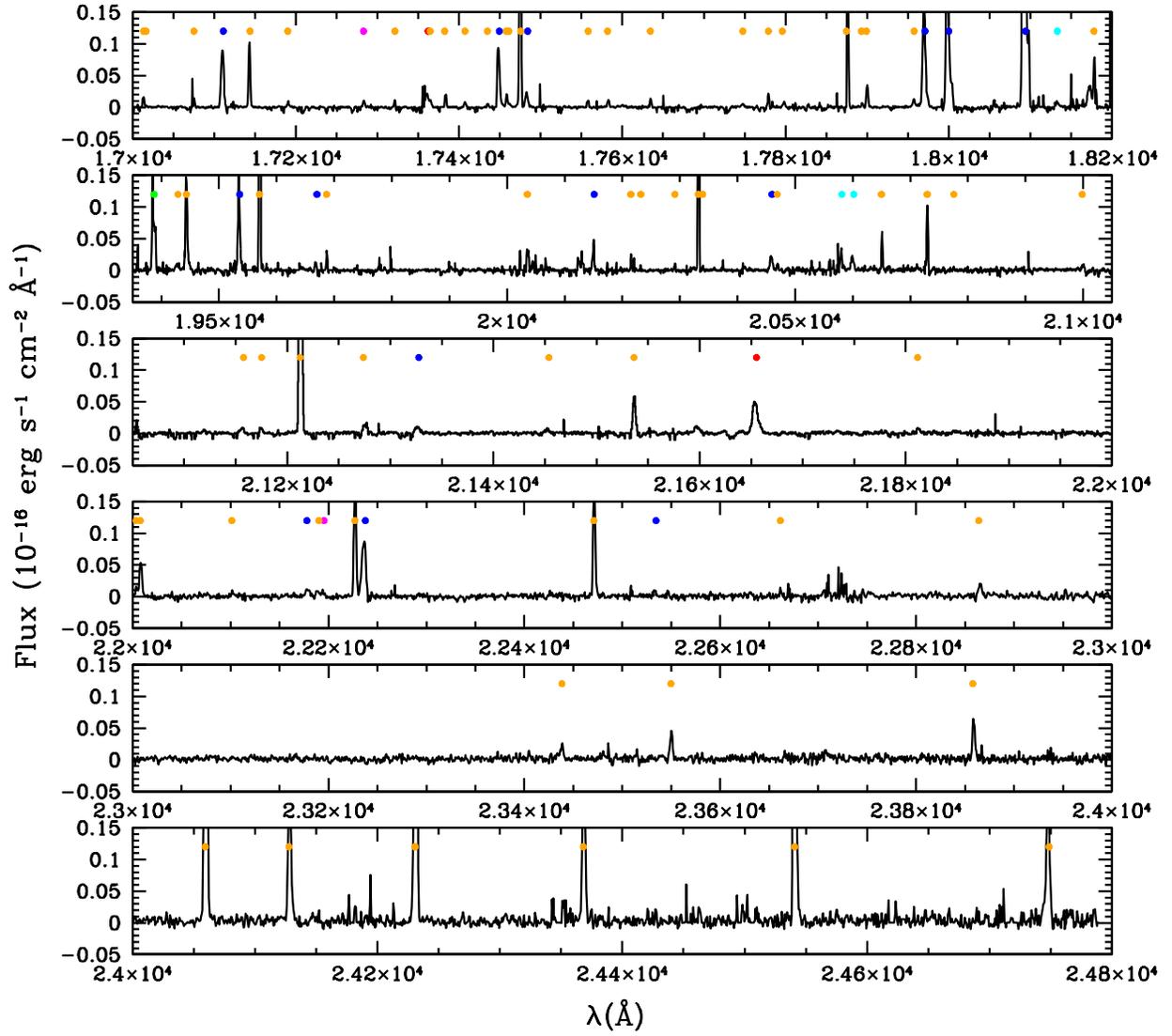}
\caption{As in Figure\,\ref{fig:fig3} in the range 17000$-$24800 \AA. \label{fig:fig5}}
\end{figure}

\begin{figure}
\epsscale{1.02}
\plotone{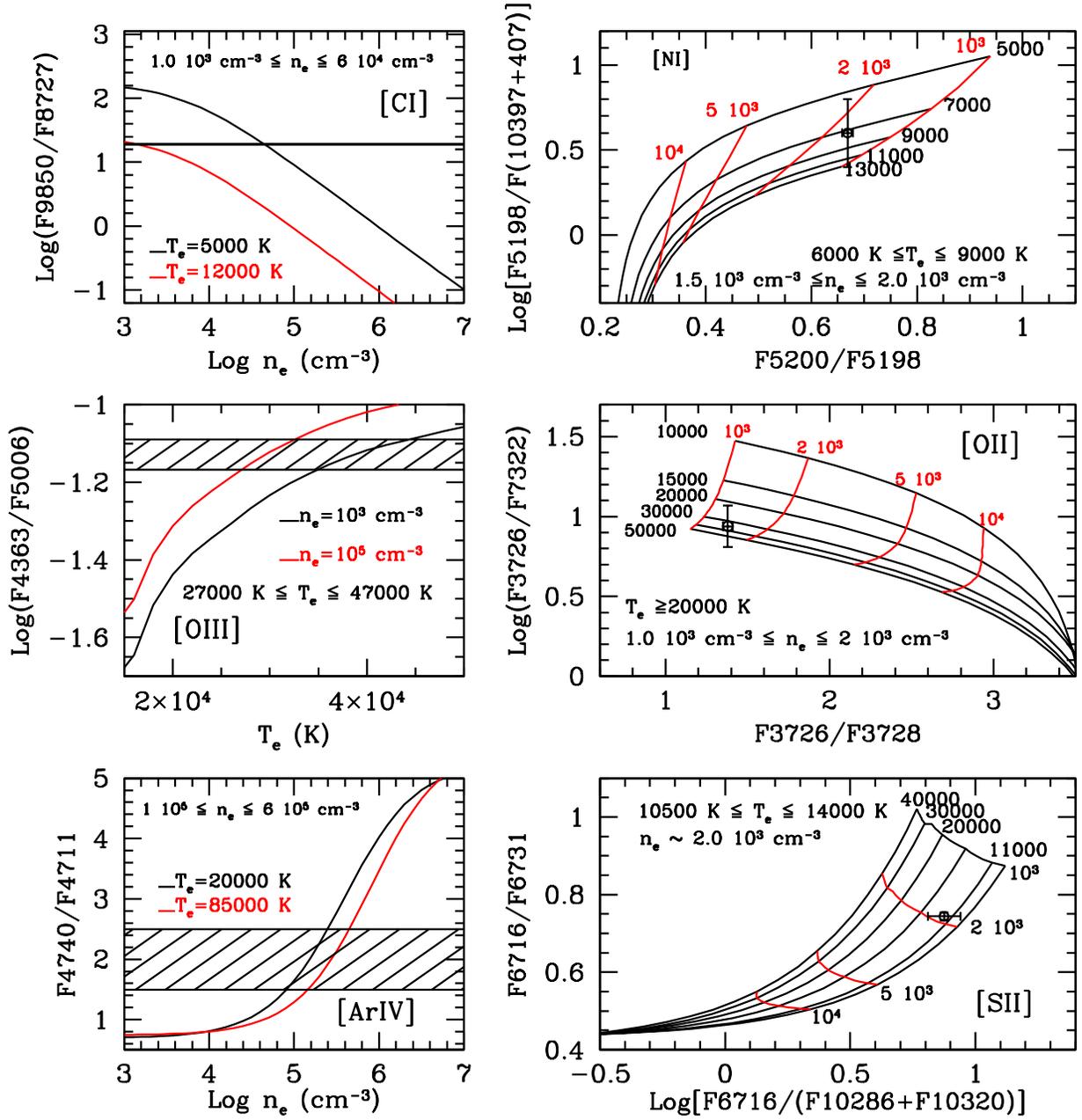}
\caption{Examples of diagnostic diagrams of atomic species where flux ratios of bright lines are plotted for different value of $T_{\rm{e}}$ and $n_{\rm{e}}$. Hatched areas in the left panels and data points in the right panels are the de-reddened flux ratios measured 
on HH\,1, considering as errors both the $\Delta$flux given in Table\,\ref{tab:tab5} and the uncertainty on A$_V$. The uncertainty due to the intercalibration between NIR and VIS arms is also considered in the ratio F5198/(F(10397+1407) plotted in the top-right panel. 
The values of $T_{\rm{e}}$ and $n_{\rm{e}}$ derived in each diagram are reported. 
\label{fig:fig6}}
\end{figure}

\begin{figure}
\epsscale{1.02}
\plotone{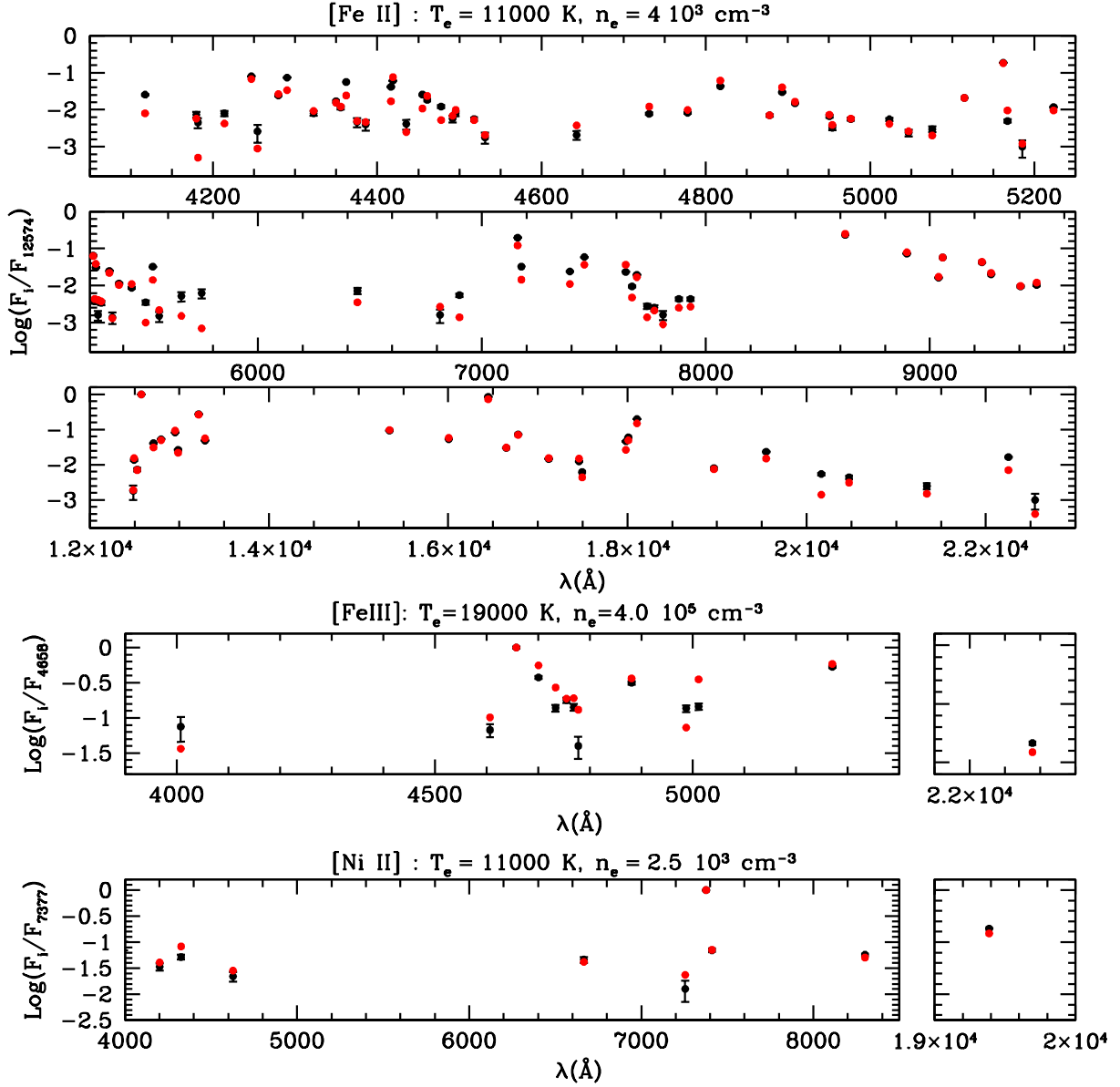}
\caption{NLTE bestfit model of the [\feii] (top three panels),  [\feiii] (fourth panel), and  [\NIii] (bottom panel). In the
fitting procedure only lines detected with snr $\ge$ 5 have been included (black: data, red: model). The best fit parameters are
reported as well. \label{fig:fig7}}
\end{figure}

\begin{figure}
\plotone{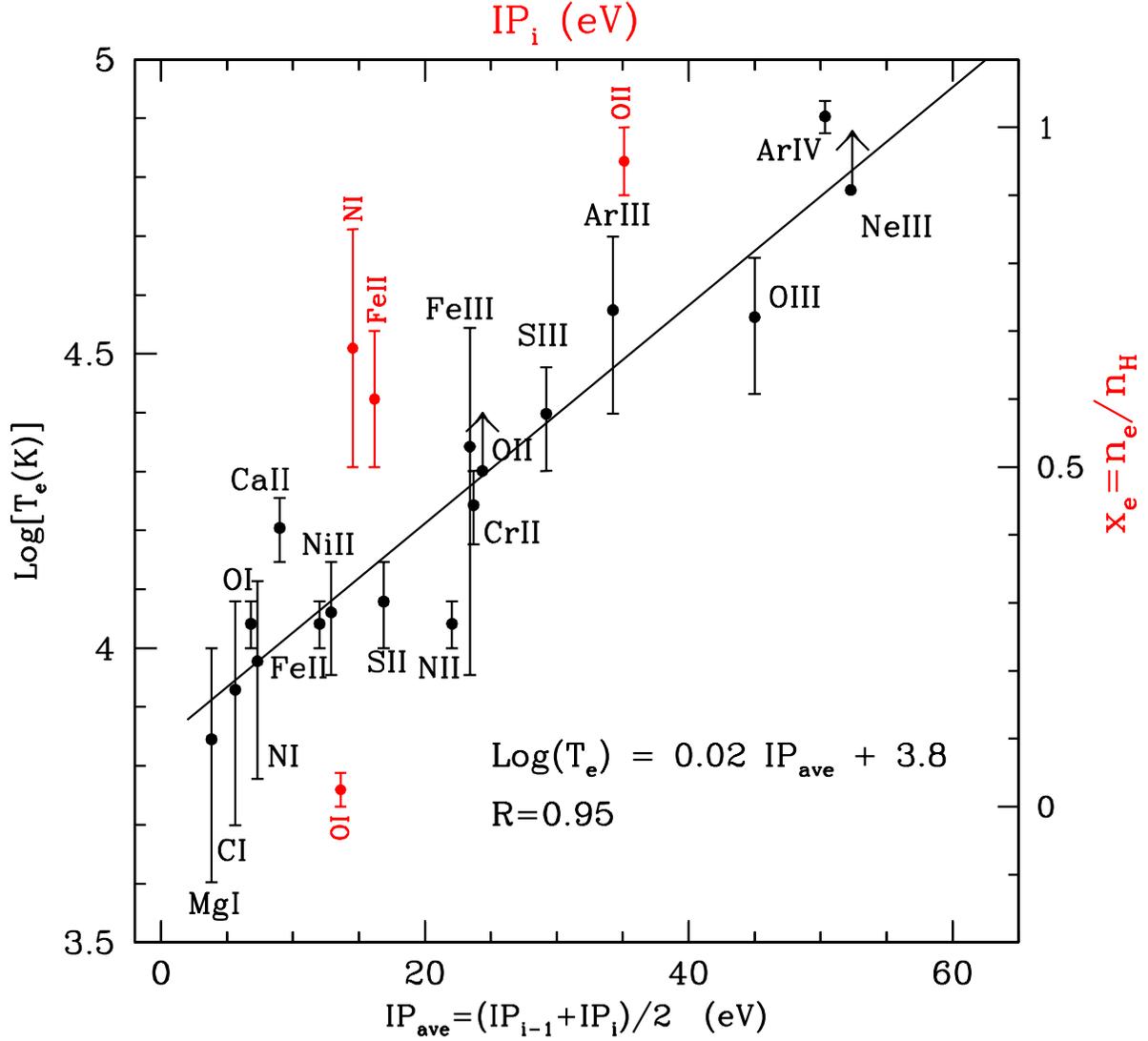}
\caption{Black: Log($T_{\rm{e}}$) derived from the excitation model vs. IP$_{ave}$. The vertical lines are the temperature uncertainties reported in Table\,\ref{tab:tab1}. The straight line is the best linear fit through the data points. The found law, along with the associated regression coefficient, is reported. Red: fractional ionization $x_{\rm{e}}$ derived from the ionization model vs. IP$_{i}$ \label{fig:fig8}}
\end{figure}

\begin{figure}
\epsscale{1.02}
\plotone{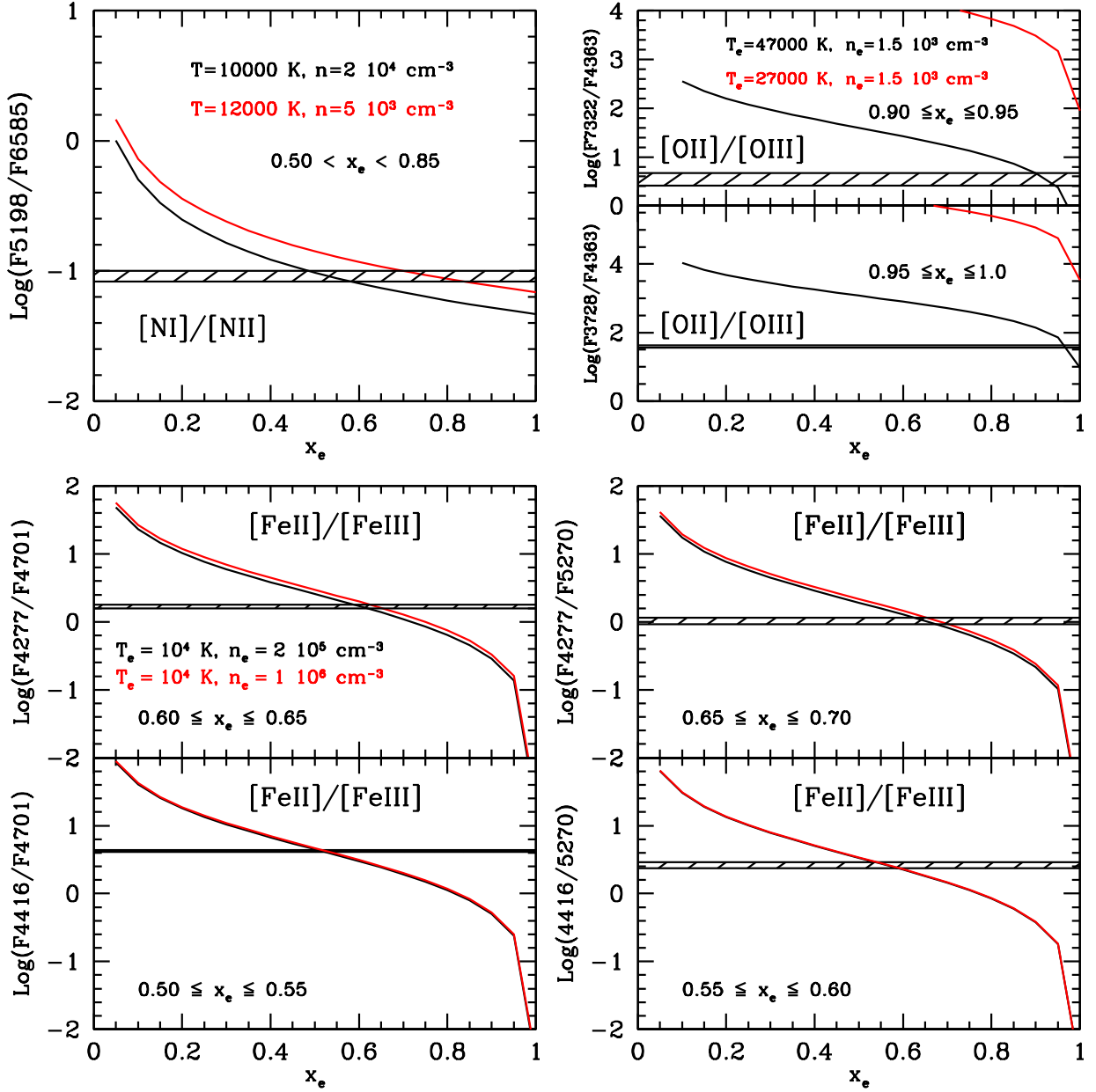}
\caption{Flux ratios of lines from adjacent ionization stages as a function of the fractional ionization $x_{\rm{e}}$. The ratios are 
plotted assuming the physical conditions reported in Table\,\ref{tab:tab1}. The hatched areas are the de-reddened flux ratios measured 
on HH\,1, considering as errors the $\Delta$flux given in Table\,\ref{tab:tab1} and the uncertainty on A$_V$. \label{fig:fig9}}
\end{figure}

\begin{figure*}
\epsscale{1.02}
\plotone{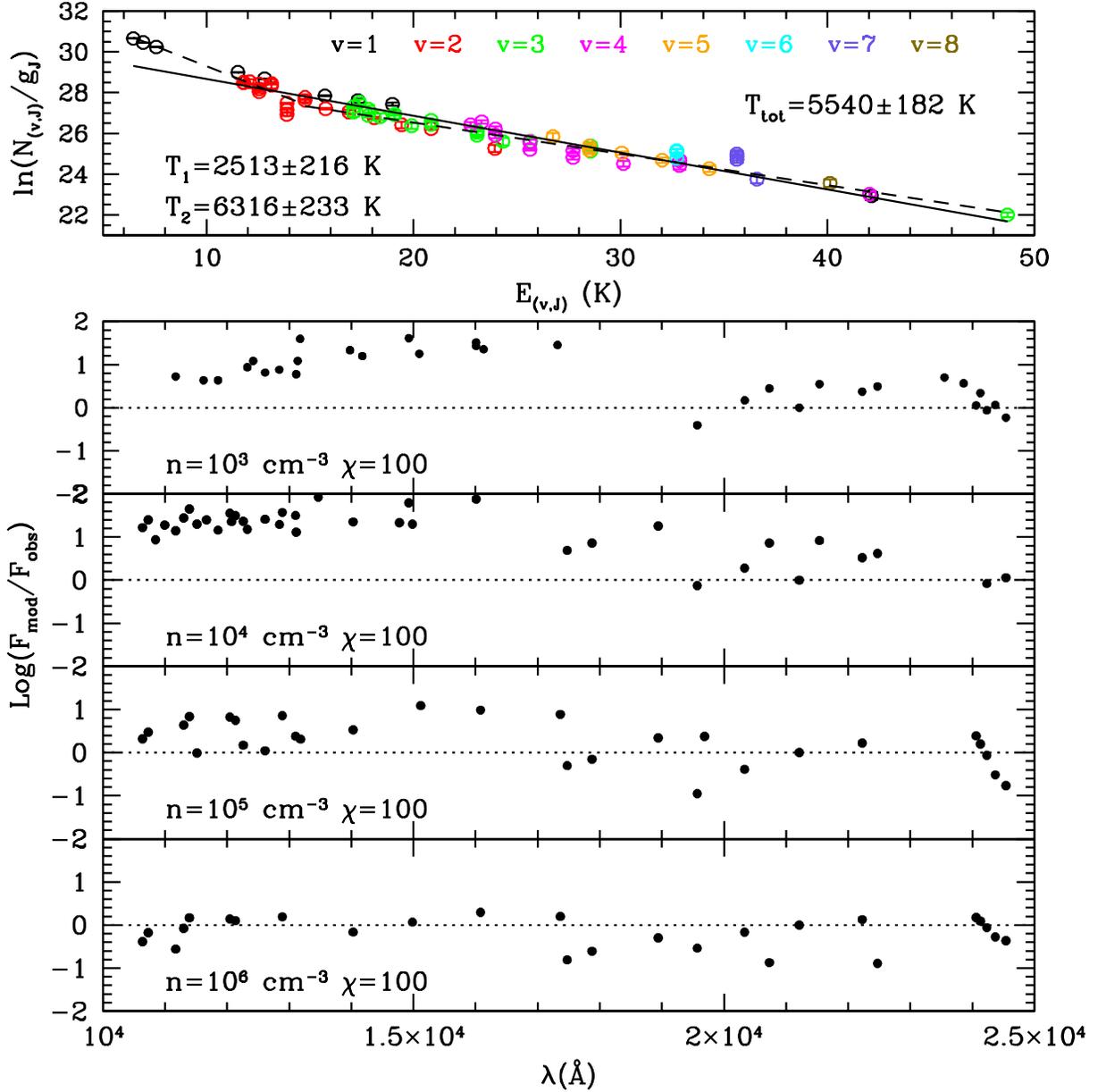}
\caption{Top: Boltzmann diagram of H$_2$ lines detected with snr $\ge$ 5. Different colors indicate lines from different vibrational states. The continuos line is the
linear best fit through all the lines (T$_{tot}$), while the dashed lines shows the fit through the $v$=1,2 lines with v$_{up}$ $\le$ 15\,000 K (T$_1$) and through all the other lines
with   v$_{up}$ $>$ 15\,000 K (T$_2$).  Bottom: comparison between the observations and the fluorescence model by Stenberg \& Dalgarno (1989). The comparison is done by assuming the density in the range 10$^3$ cm$^{-3}$ $\le n \le$ 10$^6$ cm$^{-3}$ ($n=n_{\rm{H_2}}$ + $n_{\rm{H}}$) and a radiation field strength ($\chi$) a factor of 100 higher than of the interstellar field.
 \label{fig:fig10}}
\end{figure*}

\clearpage
\begin{deluxetable}{ccccccc}
\tabletypesize{\footnotesize}
\tablecaption{\label{tab:tab1} Parameters of the excitation model.}
\tablewidth{0pt}
\tablehead{
\colhead{Ion} & \colhead{IP$_{ave}^a$} & \colhead{N$_{lev}^b$} & \colhead{Coll.rates} 
& \colhead{Range$^c$} & \colhead{$T_{\rm{e}}$} &
\colhead{$n_{\rm{e}}$}
}
\startdata
              &  (eV)          &            &                &    (10$^4$ K)    &   (10$^4$ K)   &  (10$^4$ cm$^{-3}$)  \\
\cline{1-7}
 \mgi         & 3.822          &    8       &    [1]         &  0.1   - 1       & 0.4-  1.0 (d)  &  0.1 - 1.5  \\  
 \ci          & 5.628          &    5       &    [2]         &  0.05  - 2       & 0.5 - 1.6 (e)  &  0.1 - 6.0  \\
 \oi          & 6.807          &    5       &    [3]         &  0.5   - 3       & 1.0 - 1.2 (e)  &  0.1 - 10   \\
 \n           & 7.265          &    5       &    [4]         &  0.05  - 2       & 0.6 - 1.3      &  0.15 - 0.2 \\
 \caii        & 8.989          &    5       &    [5]         &  0.3   - 3.8     & 1.4 - 1.8      &  $\le$ 100  \\
 \tii         &10.195          &   30       &    [6]         &  0.4   - 2       &   -            &  -          \\
 \feii        &12.025          &  159       &    [7]         &  0.1   - 2       & 1.0 - 1.2      &  0.3 - 0.5  \\
 \NIii        &12.891          &   17       &    [8]         &  0.2   - 3       & 0.9 - 1.4      &  0.2 - 1.0  \\
 \pii         &15.102          &    5       &    [9]         &  0.1   - 4       &     -          &   -         \\
 \sii         &16.878          &    5       &    [10]        &  0.5   - 10      & 1.0 - 1.4      &  $\sim$ 0.2 \\
 \clii        &18.405          &    5       &    [11]        &  1               &     -          &   -         \\
 \nii         &22.061          &    5       &    [12]        &  0.5   - 10      & 1.0 - 1.2 (e)  &  0.5 - 2.0  \\
 \feiii       &23.411          &   34       &    [13]        &  0.5   - 2       & 0.9 - 3.5      &  20 - 100   \\
 \crii        &23.70           &   13       &    [14]        &  0.2   - 10      & 1.5   - 2 (d)  &  0.1 - 10   \\       
 \oii         &24.631          &    5       &    [15]        &  0.5   - 2       & $\ge$ 2.0      &  0.1 - 0.2  \\
 \siii        &29.200          &    5       &    [16]        &  0.1   - 10      & 2.0 - 3.0      &  0.1 - 10   \\
 \ariii       &37.580          &    5       &    [16]        &  0.1   - 10      & 2.5 - 5.0      &  $\le$ 100  \\
 \oiii        &45.002          &    5       &    [17]        &  0.1 - 10        & 2.7 - 4.7 (e)  &  0.1 - 10   \\
 \ariv        &50.345          &    5       &    [18]        &  0.2 - 5         & 7.5 - 8.5 (f)  &  10 - 60    \\
 \neiii       &52.285          &    5       &    [19]        &  0.1 - 10        & $\ge$ 6.0 (f)  &    -        \\
\enddata
\tablenotetext{a} {IP$_{ave}$ is the average between the ionization potentials of the ionic stage {\it i}-1 and {\it i}.}
\tablenotetext{b} {Number of levels considered in the excitation model.}
\tablenotetext{c} {Range of validity of the collisional rates.}
\tablenotetext{e} {Minimum $T_{\rm{e}}$ corresponds to maximum $n_{\rm{e}}$ and viceversa.}
\tablenotetext{d} {Minimum $T_{\rm{e}}$ corresponds to minimum $n_{\rm{e}}$ and viceversa.}
\tablenotetext{f} {Temperature derived assuming X$^i$ $\le$ X$^i_\odot$.}
\tablerefs{[1]: Osorio et al. 2015; [2]: Hollenbach \& McKee, 1989; [3] Bhatia \& Kastner 1995; [4] Pequignot \& Aldrovandi, 1976; [5] Mel{\'e}ndez et al. 2007; 
[6]: Bautista et al. 2006; [7] Bautista \& Pradhan 1998;  [8] Bautista 2004; [9] Tayal 2004; [10] Tayal \& Zatsarinny 2010;
[11] Mendoza 1983; [12] Stafford et al. 1994; [13] Zhang 1996; [14] Wasson et  al. 2010; [15] Pradhan 1976;[16] Galavis et al. 1995;
[17] Lennon \& Burke 1994; [18] Zeippen et al. 1987; [19] Butler \& Zeippen 1994. }
\end{deluxetable}



\begin{deluxetable}{cccc}
\tablecaption{\label{tab:tab2} Line ratios sensitive to fluorescence.}
\tabletypesize{\footnotesize}
\tablewidth{0pt}
\tablehead{
\colhead{Ratio} & \colhead{Observed} & \colhead{Collisional} & \colhead{Coll.+ Fluor.} 
}
\startdata
\multicolumn{4}{c}{[\NIii]$^a$}\\
\cline{1-4}
7377/7411          & 0.07    & 0.05-0.07         &  0.34          \\
\cline{1-4}
\multicolumn{4}{c}{[\feii]$^b$}\\
\cline{1-4}
4815/8892         & 0.4     & 0.4               &  0.5-4          \\
4815/8617         & 0.14    & $\sim$0.17        &  0.2-0.7        \\
5334/8617         & 0.08    & 0.05-0.2          &  0.15-0.2       \\
5334/8892         & 0.25    & 0.1-0.3           &  0.5-1.2        \\
(7542+7155)/8617  & 0.9     & 0.6-1.0           &  0.3-1.0        \\
\enddata
\tablenotetext{a}{Theoretical predictions taken from Bautista et al. (1996).}
\tablenotetext{b}{Theoretical predictions taken from Bautista \& Pradhan (1998).}
\end{deluxetable}

\begin{deluxetable}{ccccc}
\tablecaption{\label{tab:tab3} Parameters of the ionization model.}
\tabletypesize{\footnotesize}
\tablewidth{0pt}
\tablehead{
\colhead{Element} & \colhead{Ionization stage} &  \colhead{$x_{\rm{e}}$ }& \colhead{$T_{\rm{e}}$} & \colhead{Fractional abundance$^{a}$} 
}
\startdata
 Helium         &  He$^{0}$/He$^{+1}$                         &  1.0$^b$       & 2.0 10$^4$                 & He$^{0}$:$\sim$100                                         \\
                &  					                          &  1.0$^b$       & 3.0 10$^4$                 & He$^{0}$ : He$^{+1}$ = 75 : 25                             \\
                &  					     	                  &  1.0$^b$       & 4.0 10$^4$                 & He$^{0}$ : He$^{+1}$ = 23 : 77                             \\
                &  				                              &  1.0$^b$       & 5.0 10$^4$                 & He$^{0}$ : He$^{+1}$ = 7 : 93                              \\
 Carbon         &  C$^{0}$/C$^{+1}$                           &  $\sim$0$^b$   & 1.0 10$^4$                 & C$^{0}$ : C$^{+1}$ = 95 : 5                                \\
 Nitrogen       &  N$^{0}$/N$^{+1}$                           &  0.50-0.85     & 1.1 10$^4$                 & N$^{0}$ : N$^{+1}$ = (53-47) : (40-60)                     \\ 
 Oxygen         &  O$^{0}$/O$^{+1}$/O$^{+2}$                  &  $\sim$0$^b$   & 1.1 10$^4$                 & O$^{0}$ : O$^{+1}$ = 99 : 1                                \\
                &                                             &  1.0           & 3.0 10$^4$                 & O$^{+1}$ : O$^{+2}$ = 99 : 1                               \\                
                &                                             &  1.0           & 4.5 10$^4$                 & O$^{+1}$ : O$^{+2}$ = 92 : 8                               \\
 Magnesium      &  Mg$^{0}$/Mg$^{+1}$                         &  $\sim$0$^b$   & 7.0 10$^3$                 & Mg$^{0}$ : Mg$^{+1}$ = 46 : 54                             \\                
 Sulphur        &  S$^{0}$/S$^{+1}$/ S$^{+2}$                 &  0.50-0.70$^b$ & 1.2 10$^4$                 & S$^{0}$ : S$^{+1}$ = 41 : 59                               \\
                &                                             &  0.50-0.70$^b$ & 3.0 10$^4$                 & S$^{0}$ : S$^{+1}$ : S$^{+2}$ = (13-6) : (72-80) :(14-16)  \\ 
 Argon          &  Ar$^{+2}$/A$^{+3}$                         &  1.0$^b$       & 3.0 10$^4$                 & Ar$^{+2}$ = 100                                            \\
                &  					                          &  1.0$^b$       & 8.0 10$^4$                 & Ar$^{+2}$ : A$^{+3}$ = 79 : 21                             \\
 Calcium        &  Na$^{0}$/Na$^{+1}$/Na$^{+2}$               &  0.50-0.85$^b$ & 1.4 10$^4$                 & Na$^{0}$ : Na$^{+}$ = 10 : 90                              \\
 Iron           &  Fe$^{0}$/Fe$^{+1}$/Fe$^{+2}$/Fe$^{+3}$     &  0.50-0.70     & 1.1 10$^4$                 & Fe$^{0}$ : Fe$^{+1}$ : Fe$^{+2}$ = (6-5) : (84-75) :(9-19) \\
                &                                             &  0.50-0.70     & 2.2 10$^4$                 & Fe$^{+1}$ : Fe$^{+2}$ = (70-50) : (30-50)                  \\
                &                                             &  0.50-0.70     & 3.5 10$^4$                 & Fe$^{+1}$ : Fe$^{+2}$ = (56-35) : (44-65)                  \\
 Nichel         &  Ni$^{0}$/Ni$^{+1}$                         &  0.50-0.85$^b$ & 2.5 10$^4$                 & Ni$^{0}$ : Ni$^{+1}$ = (86-52) : (14-48)                   \\
\enddata
\tablenotetext{a}{We report only fractional abundances $\ge$ 1\%. When a range is given, this refers to two values of $x_{\rm{e}}$ of the third column.}
\tablenotetext{b}{Assumed value (see text).}
\end{deluxetable}

\begin{deluxetable}{cccccc}
\tablecaption{\label{tab:tab4} Chemical abundances$^a$.}
\tabletypesize{\footnotesize}
\tablewidth{0pt}
\tablehead{
\colhead{Element } & \colhead{HH\,1}  & \colhead{Sun$^b$ }  & \colhead{Sun-HH\,1}   & \colhead{Orion nebula$^c$} & \colhead{Orion nebula-HH\,1}
}
\startdata
 C                 &   7.52 $\div$ 7.95     & 8.39 (0.05)    &    0.39 $\div$ 0.92  &  8.40 $\div$ 8.44          &   0.45 $\div$ 0.92          \\
 N                 &   7.57 $\div$ 7.70     & 7.78 (0.06)    &    0.02 $\div$ 0.27  &  7.65 $\div$ 7.73          &  -0.04 $\div$ 0.16          \\
 O                 &   8.60 $\div$ 8.71     & 8.66 (0.05)    &   -0.10 $\div$ 0.11  &  8.51 $\div$ 8.65          &  -0.20 $\div$ 0.04          \\
 Mg                &   6.11 $\div$ 6.77     & 7.53 (0.09)    &    0.67 $\div$ 1.51  &      -                     &      -                      \\
 P                 &   5.00 $\div$ 5.15     & 5.36 (0.04)    &    0.17 $\div$ 0.40  &      -                     &      -                      \\
 S                 &   6.78 $\div$ 7.02     & 7.14 (0.05)    &    0.07 $\div$ 0.41  &  7.06 $\div$ 7.22          &   0.04 $\div$ 0.44          \\
 Cl                &   4.97 $\div$ 5.39     & 5.50 (0.30)    &   -0.19 $\div$ 0.83  &  5.33 $\div$ 5.46          &  -0.05 $\div$ 0.49          \\
 Ar                &   6.06 $\div$ 6.23     & 6.18 (0.08)    &   -0.13 $\div$ 0.20  &  6.50 $\div$ 6.62          &   0.27 $\div$ 0.56          \\
 Ca                &   5.12 $\div$ 5.81     & 6.31 (0.04)    &    0.46 $\div$ 1.23  &      -                     &      -                      \\
 Ti                &   4.41 $\div$ 4.67     & 4.90 (0.06)    &    0.17 $\div$ 0.55  &      -                     &      -                      \\
 Cr                &   5.49 $\div$ 5.65     & 5.64 (0.10)    &   -0.11 $\div$ 0.25  &      -                     &      -                      \\ 
 Fe                &   6.91 $\div$ 7.24     & 7.45 (0.05)    &    0.16 $\div$ 0.59  & 5.86 $\div$ 6.23           &  -0.68 $\div$ -1.38         \\
 Ni                &   5.60 $\div$ 6.07     & 6.23 (0.04)    &    0.12 $\div$ 0.67  &      -                     &      -                      \\  
\enddata
\tablenotetext{a}{In units of 12+log(X/H).}
\tablenotetext{b}{From Asplund (2005).}
\tablenotetext{c}{From Esteban et al. (2004).}
\end{deluxetable}

\clearpage
\appendix
\section{Appendix}{\label{sec:appendix}}
We give here the list of the detected lines. Table\,\ref{tab:tab5}: atomic forbidden lines, 
Table\,\ref{tab:tab6}: Hydrogen and Helium recombination lines, Table\,\ref{tab:tab7} : \htwo\, ro-vibrational lines.

\begin{deluxetable}{cccccccc}
\tablewidth{0pt}
\tabletypesize{\footnotesize}
\tablecaption{Atomic forbidden lines \label{tab:tab5}}
\tablehead{
\colhead{$\lambda _{air}$}           & \colhead{Ion}  &
\colhead{Upp.lev.}          & \colhead{Low. lev.}  &
\colhead{$E_{up}$}       & \colhead{$E_{lo}$}    &
\colhead{Flux$^{a,b}$}    &
\colhead{$\Delta$(Flux)} 
}
\startdata
(\AA) & &    &  & (cm$^{-1}$) & (cm$^{-1}$)& \multicolumn{2}{c}{(10$^{-17}$\,erg\,s$^{-1}$\,cm$^{-2}$)}  \\
\cline{1-8}\\[-5pt]

\multicolumn{8}{c}{Carbon $-$ Z=6} \\ 

\cline{1-8}\\[-5pt]

8727.12 & [\ci]   & $^1\!S_{0}$ & $^1\!D_{2}$  &	21 648.0 &  10 192.63   & 13.5 + 3.9  & 0.5 \\

9824.13 & [\ci]   & $^1\!D_{2}$ & $^3\!P_{1}$  & 	10 192.63 & 16.40   	& 88.8 + 21.7  & 0.5\\

9850.26 & [\ci]   & $^1\!D_{2}$ & $^3\!P_{2}$  & 	10 192.63 & 43.40   	& 262.8 + 66.4  & 0.5  \\

\cline{1-8}\\[-5pt]

\multicolumn{8}{c}{Nitrogen $-$ Z=7} \\  

\cline{1-8}\\[-5pt]

5197.90 & [\n] & $^2\!Do_{3/2}$ & $^4\!So_{3/2}$ & 19 233.18  & 0.00   	& 84.3 + 10.8  & 0.3  \\

5200.26 & [\n] & $^2\!Do_{5/2}$ & $^4\!So_{3/2}$ & 19 224.46   & 0.00     & 57.3 + 6.3  & 0.3  \\

10397.74& [\n] & $^2\!Po_{3/2}$ & $^2\!Do_{5/2}$ & 28 839.31   & 19 224.46 & 17.2     & 0.7\\ 

10398.15& [\n] & $^2\!Po_{1/2}$ & $^2\!Do_{5/2}$ & 28 838.92   & 19 224.46 & 5.6        & 0.7 \\ 

10407.17& [\n] & $^2\!Po_{3/2}$ & $^2\!Do_{3/2}$ & 28 839.31   & 19 233.17& 10.8      & 0.7   \\

10407.59& [\n] & $^2\!Po_{1/2}$ & $^2\!Do_{3/2}$ & 28 838.92   & 19 233.17 & 1.8       & 0.7  \\

5754.64 &[\nii]& $^1\!S_{0}$    & $^1\!D_{2}$    & 32 688.64   & 15 316.2  & 25.3 + 6.3  & 0.7 \\
  
6548.04 &[\nii]& $^1\!D_{2}$    & $^3\!P_{1}$    & 15 316.19   & 48.7 & 329.0 + 63.9  & 0.6\\

6583.46 &[\nii]& $^1\!D_{2}$    & $^3\!P_{2}$    & 15 316.19   & 130.8      &1020.0 + 207.2 & 1.1\\

\cline{1-8}
\enddata
\tablenotetext{a}{We do not report the flux of lines well detected both in the spectral images and the 1-D spectrum but contaminated by image artifacts or located in spectral segment with poor atmospheric transmission.}
\tablenotetext{b}{If the line presents two flux components these are reported as F$_1$+F$_2$.}
\tablecomments{
1-Blended with [\oii] $^2\!Po_{3/2} - ^2\!Do_{5/2}$;
2-Blended with [\oii] $^2\!Po_{1/2} - ^2\!Do_{5/2}$;
3-Blended with [\oii] $^2\!Po_{3/2} - ^2\!Do_{3/2}$;
4-Blended with [\oii] $^2\!Po_{1/2} - ^2\!Do_{3/2}$;
5-Blended with [\feii] $b^4\!P_{5/2} - a^6\!D_{5/2}$;
6-Blended with [\feii] $b^4\!P_{5/2} - a^6\!D_{3/2}$;
7-Blended with [\feiii] $^3\!G_{4} -^5\!D_{4}$;
8-Blended with [\feii] $a^2\!F_{7/2} - a^4\!P_{3/2}$;
9-Blended with  [\crii]  $b^4\!D_{7/2} - a^6\!S_{5/2}$ and [\NIii]  $^2\!D_{5/2} - ^2\!D_{5/2}$;
10-Blended with  [\crii]  $b^4\!D_{1/2} - a^6\!S_{5/2}$ and [\NIii]  $^2\!D_{5/2} - ^2\!D_{5/2}$;
11-Blended with  [\crii] $a^4\!P_{5/2} - a^6\!S_{5/2}$;
12-Blended with  [\crii] $a^4\!P_{3/2} - a^6\!S_{5/2}$;
13-Blended with  [\feii] $a^4\!P_{5/2} - a^6\!D_{1/2}$;
14-Blended with  [\feii] $b^2\!P_{3/2} - a^6\!D_{7/2}$;
15-Blended with  [\feii] $a^4\!G_{9/2} - a^6\!D_{7/2}$;
16-Blended with  [\feii] $a^2\!F_{5/2} - a^4\!F_{7/2}$;
17-Blended with  [\feii] $a^4\!G_{5/2} - a^6\!D_{3/2}$;
18-Blended with [\feii] $a^4\!G_{7/2} - a^4\!F_{7/2}$;
19-Blended with [\feii] $a^4\!G_{11/2} - a^4\!F_{9/2}$;
20-Blended with [\feii] $b^4\!F_{3/2} - a^6\!D_{7/2}$;
21-Blended with [\feii] $a^6\!S_{5/2} - a^6\!D_{5/2}$;
22-Blended with \he\ $^3\!D - ^3\!Po$;
23-Blended with  [\oiii] $^1\!D_2 - ^3\!P_1$;
24-Blended with  [\oiii] $^1\!D_2 - ^3\!P_2$;
25-Blended with  [\feiii] $^3\!P_{4} - ^5\!D_{1}$;
26-Blended with  [\caii] $^2\!D_{3/2} - ^2\!S_{1/2}$;
27-Blended with  [\feii]  $a^4\!P_{3/2} - a^6\!D_{5/2}$;
28-Blended with  [\feii] $a^2\!G_{9/2} - a^4\!F_{5/2}$;
29-Blended with \htwo\, 3-0 S(13);
30-Blended with [\crii] $a^6\!D_{9/2} - a^6\!S_{5/2}$;
31-Blended with \he\, $^3\!D $-$ ^3\!Po$;
32-Blended with [\feii] $b^2\!H_{11/2} - a^2\!H_{11/2}$ and \htwo\, 3-1 O(8);
33-Blended with [\feii] $b^2\!F_{5/2} - a^4\!G_{7/2}$ and \htwo\, 3-1 O(8);
34-Blended with [\sii] $^2\!Po_{3/2} - ^4\!So_{3/2}$;
35-Blended with [\feii] $a^2\!D2_{3/2} - a^4\!F_{5/2}$;
36-Blended with \htwo\, 7-5 Q(1);
37-Blended with [\crii]  $b^4\!D_{1/2} - a^6\!S_{5/2}$ and [\crii]  $b^4\!D_{7/2} - a^6\!S_{5/2}$;
38-Blended with \htwo\, 4-1 S(6).
}
\tablecomments{Table 5 is published in its entirety in the electronic 
edition of the {\it Astrophysical Journal}.  A portion is shown here 
for guidance regarding its form and content.}

\end{deluxetable}

\begin{deluxetable}{rcccccccc}
\tablewidth{0pt}
\tabletypesize{\footnotesize}
\tablecaption{Recombination lines \label{tab:tab6}}
\tablehead{
\colhead{$\lambda _{air}$}           & 
\colhead{Upper}          & \colhead{Lower}  &
\colhead{$E_{up}$}       & \colhead{$E_{lo}$}    &
\colhead{Flux$^a$}   &\colhead{$\Delta$(Flux)}
}
\startdata
(\AA) &  level & level & (cm$^{-1}$) & (cm$^{-1}$)& \multicolumn{2}{c}{(10$^{-17}$\,erg\,s$^{-1}$\,cm$^{-2}$)}  \\
\cline{1-7}\\[-5pt]
\multicolumn{7}{c}{Hydrogen} \\
\cline{1-7}\\[-5pt]
3669.46 & 25 & 2 & 109 503.29 &  82 259.11 & 1.6  &  0.9  \\
3671.47 & 24 & 2 & 109 488.36 &  82 259.11&  2.6  &  0.9  \\
3673.76 & 23 & 2 & 109 471.44 &  82 259.11&  1.9  &  0.9  \\
3676.36 & 22 & 2 & 109 452.16 &  82 259.11&  1.8  &  0.9  \\
3679.35 & 21 & 2 & 109 430.07 &  82 259.11&  3.2  &  0.9 \\
\enddata
\tablenotetext{a}{If the line presents two flux components these are reported as F$_1$+F$_2$.}
\tablecomments{
1-Blended with \htwo\, 7-5 S(2);
2-Blended with [\feii] $b^4\!F_{3/2} - a^6\!D_{5/2}$;
3-Blended with [\feii] $a^2\!D2_{5/2}-a^4\!D_{3/2}$.
}
\tablecomments{Table 6 is published in its entirety in the electronic 
edition of the {\it Astrophysical Journal}.  A portion is shown here 
for guidance regarding its form and content.}
\end{deluxetable}

\begin{deluxetable}{rccccccc}
\tablewidth{0pt}
\tabletypesize{\footnotesize}
\tablecaption{\htwo\, lines \label{tab:tab7}}
\tablehead{
\colhead{$\lambda _{air}$}           & 
\colhead{Transition}           &
\colhead{$E_{up}$}       & \colhead{$E_{lo}$}    &
\colhead{Flux$^a$}   &\colhead{$\Delta$(Flux)} 
}
\startdata
(\AA) &   & (K) & (K)& \multicolumn{2}{c}{(10$^{-17}$\,erg\,s$^{-1}$\,cm$^{-2}$)}  \\
\cline{1-6}\\[-5pt]
\multicolumn{6}{c}{v=1} \\
\cline{1-6}\\[-5pt]
16452.56 & 1-0 S(17)& 30 156.1    &  21 412.7  & 1.0  & 0.4   \\
16499.78 & 1-0 S(11)& 18 980.6    &  10 262.3  & 5.0  & 0.4   \\
16582.11 & 1-0 S(18)& 32 136.1    &  23 461.0  & 0.7  & 0.4   \\
16660.36 & 1-0 S(10)& 17 312.0    &  8 677.73  & 3.6  & 0.4   \\
16745.69 & 1-0 S(19)& 34 131.1    &  25 540.8  & 1.2  & 0.4   \\
\enddata
\tablenotetext{a}{We do not report the flux of lines well detected both in the spectral images and the 1-D spectrum but contaminated by image artifacts or located in spectral segment with poor atmospheric transmission.}
\tablecomments {
1-Blended with \htwo\, 7-4 S(12);
2-Blended with \htwo\, 3-1 S(6);
3-Blended with  [\feii]  $b^4\!D_{5/2} - a^2\!P_{1/2}$;
4-Blended with \htwo\, 6-3 S(7);
5-Blended with \htwo\, 7-4 Q(3);
6-Blended with \htwo\, 2-0 S(19);
7-Blended with  [\feii] $b^2\!F_{5/2} - a^4\!G_{7/2}$ and [\feii] $b^2\!H_{11/2} - a^2\!H_{11/2}$;
8-Blended with [\NIii]  $^2\!F_{7/2} - ^2\!D_{3/2}$;
9-Blended with \htwo\, 6-3 S(6);
10-Blended with \htwo\, 4-2 S(6);
11-Blended with \htwo\, 4-2 S(13);
12-Blended with \htwo\, 5-3 S(7) and \htwo\, 5-3 S(11);
13-Blended with \htwo\, 5-3 S(8);
14-Blended with \htwo\, 5-3 S(10);
15-Blended with \htwo\, 5-3 S(7) and \htwo\, 4-1 O(13);
16-Blended with \htwo\, 5-3 S(11) and \htwo\, 4-1 O(13);
17-Blended with \htwo\, 3-0 Q(12);
18-Blended with \htwo\, 4-1 O(3);
19-Blended with \htwo\, 2-0 S(15);
20-Blended with \htwo\, 3-1 S(10);
21-Blended with \hi\, 14-4;
22-Blended with [\coii]  $b^3\!F_{4} - a^5\!F_{4}$.
}
\tablecomments{Table 7 is published in its entirety in the electronic 
edition of the {\it Astrophysical Journal}.  A portion is shown here 
for guidance regarding its form and content.}
\end{deluxetable}


\begin{thebibliography}{}
\bibitem[Agra-Amboage et al.(2011)]{2011A&A...532A..59A} Agra-Amboage, V., Dougados, C., Cabrit, S., \& Reunanen, J.\ 2011, \aap, 532, A59 
\bibitem[Arnaud \& Raymond(1992)]{1992ApJ...398..394A} Arnaud, M., \& Raymond, J.\ 1992, \apj, 398, 394 
\bibitem[Asplund(2005)]{2005ARA&A..43..481A} Asplund, M.\ 2005, \araa, 43, 481 
\bibitem[Bautista et al.(1996)]{1996ApJ...460..372B} Bautista, M.~A., Peng, J., \& Pradhan, A.~K.\ 1996, \apj, 460, 372 
\bibitem[Bautista(2004)]{2004A&A...420..763B} Bautista, M.~A.\ 2004, \aap, 420, 763 
\bibitem[Bautista et al.(2006)]{2006MNRAS.370.1991B} Bautista, M.~A., Hartman, H., Gull, T.~R., Smith, N., \& Lodders, K.\ 2006, \mnras, 370, 1991 
\bibitem[Bautista \& Pradhan(1998)]{1998ApJ...492..650B} Bautista, M.~A., \& Pradhan, A.~K.\ 1998, \apj, 492, 650 
\bibitem[Bhatia \& Kastner(1995)]{1995ApJS...96..325B} Bhatia, A.~K., \& Kastner, S.~O.\ 1995, \apjs, 96, 325 
\bibitem[Beck-Winchatz et al.(1994)]{1994PASP..106.1271B} Beck-Winchatz, B., Bohm, K.~H., \& Noriega-Crespo, A.\ 1994, \pasp, 106, 1271 
\bibitem[Beck-Winchatz et al.(1996)]{1996AJ....111..346B} Beck-Winchatz, B., Bohm, K.-H., \& Noriega-Crespo, A.\ 1996, \aj, 111, 346 
\bibitem[B{\"o}hm \& Matt(2001)]{2001PASP..113..158B} B{\"o}hm, K.-H., \& Matt, S.\ 2001, \pasp, 113, 158 
\bibitem[Boehm et al.(1993)]{1993ApJ...416..647B} B{\"o}hm, K.-H., Noriega-Crespo, A., \& Solf, J.\ 1993, \apj, 416, 647 
\bibitem[Biazzo et al.(2012)]{2012A&A...547A.104B} Biazzo, K., Alcal{\'a}, J.~M., Covino, E., et al.\ 2012, \aap, 547, A104 
\bibitem[Brugel et al.(1981)]{1981ApJ...243..874B} Brugel, E.~W., Boehm, K.~H., \& Mannery, E.\ 1981, \apj, 243, 874 
\bibitem[Butler \& Zeippen(1994)]{1994A&AS..108....1B} Butler, K., \& Zeippen, C.~J.\ 1994, \aaps, 108, 1 
\bibitem[Cardelli et al.(1989)]{1989ApJ...345..245C} Cardelli, J.~A., Clayton, G.~C., \& Mathis, J.~S.\ 1989, \apj, 345, 245 
\bibitem[Dopita et al.(1982)]{1982ApJ...261..183D} Dopita, M.~A., Binette, L., \& Schwartz, R.~D.\ 1982, \apj, 261, 183 
\bibitem[Draine(2003)]{2003ApJ...598.1026D} Draine, B.~T.\ 2003, \apj, 598, 1026 
\bibitem[Eisl{\"o}ffel et al.(2000)]{2000A&A...359.1147E} Eisl{\"o}ffel, J., Smith, M.~D., \& Davis, C.~J.\ 2000, \aap, 359, 1147 
\bibitem[Esteban et al.(2004)]{2004MNRAS.355..229E} Esteban, C., Peimbert, M., Garc{\'{\i}}a-Rojas, J., et al.\ 2004, \mnras, 355, 229 
\bibitem[Flower et al.(2003)]{2003MNRAS.341...70F} Flower, D.~R., Le Bourlot, J., Pineau des For{\^e}ts, G., \& Cabrit, S.\ 2003, \mnras, 341, 70 
\bibitem[Galavis et al.(1995)]{1995A&AS..111..347G} Galavis, M.~E., Mendoza, C., \& Zeippen, C.~J.\ 1995, \aaps, 111, 347 
\bibitem[Garcia Lopez et al.(2008)]{2008A&A...487.1019G} Garcia Lopez, R., Nisini, B., Giannini, T., et al.\ 2008, \aap, 487, 1019 
\bibitem[Giannini et al.(2008)]{2008A&A...481..123G} Giannini, T., Calzoletti, L., Nisini, B., et al.\ 2008, \aap, 481, 123 
\bibitem[Giannini et al.(2013)]{2013ApJ...778...71G} Giannini, T., Nisini, B., Antoniucci, S., et al.\ 2013, \apj, 778, 71 
\bibitem[Giannini et al.(2015)]{2015ApJ...798...33G} Giannini, T., Antoniucci, S., Nisini, B., et al.\ 2015, \apj, 798, 33, Paper I
\bibitem[Gredel(1994)]{1994A&A...292..580G} Gredel, R.\ 1994, \aap, 292, 580 
\bibitem[Gredel(1996)]{1996A&A...305..582G} Gredel, R.\ 1996, \aap, 305, 582 
\bibitem[Grevesse \& Sauval(1998)]{1998SSRv...85..161G} Grevesse, N., \& Sauval, A.~J.\ 1998, \ssr, 85, 161 
\bibitem[Guillet et al.(2009)]{2009A&A...497..145G} Guillet, V., Jones, A.~P., \& Pineau Des For{\^e}ts, G.\ 2009, \aap, 497, 145 
\bibitem[Jones(2000)]{2000JGR...10510257J} Jones, A.~P.\ 2000, \jgr, 105, 10257 
\bibitem[Jones et al.(1994)]{1994ApJ...433..797J} Jones, A.~P., Tielens, A.~G.~G.~M., Hollenbach, D.~J., \& McKee, C.~F.\ 1994, \apj, 433, 797 
\bibitem[Haro(1952)]{1952ApJ...115..572H} Haro, G.\ 1952, \apj, 115, 572 
\bibitem[Hartigan et al.(2011)]{2011ApJ...736...29H} Hartigan, P., Frank, A., Foster, J.~M., et al.\ 2011, \apj, 736, 29 
\bibitem[Hartigan \& Morse(2007)]{2007ApJ...660..426H} Hartigan, P., \& Morse, J.\ 2007, \apj, 660, 426 
\bibitem[Hartigan et al.(1994)]{1994ApJ...436..125H} Hartigan, P., Morse, J.~A., \& Raymond, J.\ 1994, \apj, 436, 125 
\bibitem[Herbig(1951)]{1951ApJ...113..697H} Herbig, G.~H.\ 1951, \apj, 113, 697 
\bibitem[Hollenbach \& McKee(1989)]{1989ApJ...342..306H} Hollenbach, D., \& McKee, C.~F.\ 1989, \apj, 342, 306 
\bibitem[Kingdon \& Ferland(1996)]{1996ApJS..106..205K} Kingdon, J.~B., \& Ferland, G.~J.\ 1996, \apjs, 106, 205 
\bibitem[Landini \& Monsignori Fossi(1990)]{1990A&AS...82..229L} Landini, M., \& Monsignori Fossi, B.~C.\ 1990, \aaps, 82, 229 
\bibitem[Lennon \& Burke(1994)]{1994A&AS..103..273L} Lennon, D.~J., \& Burke, V.~M.\ 1994, \aaps, 103, 273 
\bibitem[Lucy(1995)]{1995A&A...294..555L} Lucy, L.~B.\ 1995, \aap, 294, 555 
\bibitem[Maurri et al.(2014)]{2014A&A...565A.110M} Maurri, L., Bacciotti, F., Podio, L., et al.\ 2014, \aap, 565, A110 
\bibitem[May et al.(2000)]{2000MNRAS.318..809M} May, P.~W., Pineau des For{\^e}ts, G., Flower, D.~R., et al.\ 2000, \mnras, 318, 809 
\bibitem[Mel{\'e}ndez et al.(2007)]{2007A&A...469.1203M} Mel{\'e}ndez, M., Bautista, M.~A., \& Badnell, N.~R.\ 2007, \aap, 469, 1203 
\bibitem{} Mendoza, C. 1983 IAUS, 103, 143 
\bibitem[Mesa-Delgado et al.(2009)]{2009MNRAS.395..855M} Mesa-Delgado, A., Esteban, C., Garc{\'{\i}}a-Rojas, J., et al.\ 2009, \mnras, 395, 855 
\bibitem[Mewe et al.(1986)]{1986A&AS...65..511M} Mewe, R., Lemen, J.~R., \& van den Oord, G.~H.~J.\ 1986, \aaps, 65, 511 
\bibitem[Modigliani et al.(2010)]{2010SPIE.7737E..28M} Modigliani, A., Goldoni, P., Royer, F., et al.\ 2010, \procspie, 7737, 773728 
\bibitem[Mouri \& Taniguchi(2000)]{2000ApJ...534L..63M} Mouri, H., \& Taniguchi, Y.\ 2000, \apjl, 534, L63 
\bibitem[Nisini et al.(2005)]{2005A&A...441..159N} Nisini, B., Bacciotti, F., Giannini, T., et al.\ 2005, \aap, 441, 159 
\bibitem[Nisini et al.(2002)]{2002A&A...393.1035N} Nisini, B., Caratti o Garatti, A., Giannini, T., \& Lorenzetti, D.\ 2002, \aap, 393, 1035 
\bibitem[Nisini et al.(2010)]{2010ApJ...724...69N} Nisini, B., Giannini, T., Neufeld, D.~A., et al.\ 2010, \apj, 724, 69 
\bibitem[Nussbaumer \& Storey(1983)]{1983A&A...126...75N} Nussbaumer, H., \& Storey, P.~J.\ 1983, \aap, 126, 75 
\bibitem[Ortolani \& Dodorico(1980)]{1980A&A....83L...8O} Ortolani, S., \& Dodorico, S.\ 1980, \aap, 83, L8 
\bibitem[Osorio et al.(2015)]{2015A&A...579A..53O} Osorio, Y., Barklem, P.~S., Lind, K., et al.\ 2015, \aap, 579, A53
\bibitem[Pequignot \& Aldrovandi(1976)]{1976A&A....50..141P} Pequignot, D., \& Aldrovandi, S.~M.~V.\ 1976, \aap, 50, 141 
\bibitem[Pesenti et al.(2003)]{2003A&A...410..155P} Pesenti, N., Dougados, C., Cabrit, S., et al.\ 2003, \aap, 410, 155 
\bibitem[Pradhan(1976)]{1976MNRAS.177...31P} Pradhan, A.~K.\ 1976, \mnras, 177, 31 
\bibitem[Pravdo et al.(2001)]{2001Natur.413..708P} Pravdo, S.~H., Feigelson, E.~D., Garmire, G., et al.\ 2001, \nat, 413, 708 
\bibitem[Pravdo et al.(1985)]{1985ApJ...293L..35P} Pravdo, S.~H., Rodriguez, L.~F., Curiel, S., et al.\ 1985, \apjl, 293, L35 
\bibitem[Podio et al.(2006)]{2006A&A...456..189P} Podio, L., Bacciotti, F., Nisini, B., et al.\ 2006, \aap, 456, 189 
\bibitem[Podio et al.(2011)]{2011A&A...527A..13P} Podio, L., Eisl{\"o}ffel, J., Melnikov, S., Hodapp, K.~W., \& Bacciotti, F.\ 2011, \aap, 527, A13 
\bibitem[Podio et al.(2009)]{2009A&A...506..779P} Podio, L., Medves, S., Bacciotti, F., Eisl{\"o}ffel, J., \& Ray, T.\ 2009, \aap, 506, 779 
\bibitem[Quinet.(2007)] {1997Phys Sr...55..41} Quinet, P. \ 1997, Phys. Scr., 55, 41 
\bibitem[Raga et al.(2011)]{2011RMxAA..47..425R} Raga, A.~C., Reipurth, B., Cant{\'o}, J., Sierra-Flores, M.~M., \& Guzm{\'a}n, M.~V.\ 2011, \rmxaa, 47, 425 
\bibitem[Raga et al.(2015)]{2015ApJ...798L...1R} Raga, A.~C., Reipurth, B., Castellanos-Ram{\'{\i}}rez, A., Chiang, H.-F., \& Bally, J.\ 2015, \apjl, 798, L1 
\bibitem[Rodr{\'{\i}}guez et al.(2000)]{2000AJ....119..882R} Rodr{\'{\i}}guez, L.~F., Delgado-Arellano, V.~G., G{\'o}mez, Y., et al.\ 2000, \aj, 119, 882 
\bibitem[Savage \& Sembach(1996)]{1996ARA&A..34..279S} Savage, B.~D., \& Sembach, K.~R.\ 1996, \araa, 34, 279 
\bibitem[Solf et al.(1988)]{1988ApJ...334..229S} Solf, J., Bohm, K.~H., \& Raga, A.\ 1988, \apj, 334, 229 
\bibitem[Spina et al.(2014)]{2014A&A...568A...2S} Spina, L., Randich, S., Palla, F., et al.\ 2014, \aap, 568, A2 
\bibitem[Stafford et al.(1994)]{1994MNRAS.268..816S} Stafford, R.~P., Bell, K.~L., Hibbert, A., \& Wijesundera, W.~P.\ 1994, \mnras, 268, 816 
\bibitem[Stancil et al.(1998)]{1998ApJ...502.1006S} Stancil, P.~C., Havener, C.~C., Krsti{\'c}, P.~S., et al.\ 1998, \apj, 502, 1006 
\bibitem[Sternberg \& Dalgarno(1989)]{1989ApJ...338..197S} Sternberg, A., \& Dalgarno, A.\ 1989, \apj, 338, 197 
\bibitem[Storey \& Hummer(1995)]{1995MNRAS.272...41S} Storey, P.~J., \& Hummer, D.~G.\ 1995, \mnras, 272, 41 
\bibitem[Tayal(2004)]{2004ApJS..150..465T} Tayal, S.~S.\ 2004, \apjs, 150, 465 
\bibitem[Tayal \& Zatsarinny(2010)]{2010ApJS..188...32T} Tayal, S.~S., \& Zatsarinny, O.\ 2010, \apjs, 188, 32 
\bibitem[Vernet et al.(2011)]{2011A&A...536A.105V} Vernet, J., Dekker, H., D'Odorico, S., et al.\ 2011, \aap, 536, AA105 
\bibitem[Wasson et al.(2010)]{2010A&A...524A..35W} Wasson, I.~R., Ramsbottom, C.~A., \& Norrington, P.~H.\ 2010, \aap, 524, A35 
\bibitem[Wilson \& Rood(1994)]{1994ARA&A..32..191W} Wilson, T.~L., \& Rood, R.\ 1994, \araa, 32, 191 
\bibitem[Zeippen et al.(1987)]{1987A&A...188..251Z} Zeippen, C.~J., Butler, K., \& Le Bourlot, J.\ 1987, \aap, 188, 251 
\bibitem[Zhang(1996)]{1996A&AS..119..523Z} Zhang, H.\ 1996, \aaps, 119, 523 


\end{thebibliography}
\end{document}